\documentclass{aa}  

\usepackage{graphicx}
\usepackage{txfonts}
\usepackage{rotating}
\usepackage{subfig}
\usepackage{lscape}
\usepackage{caption}
\usepackage{amsmath}

\usepackage{color}
\usepackage{soul}
\usepackage[dvipsnames]{xcolor}
\newcommand{\unit}[1]{\,\mathrm{#1}}

\usepackage{natbib,twoopt}
\usepackage[breaklinks=true]{hyperref} 
\bibpunct{(}{)}{;}{a}{}{,}             
\makeatletter
  \newcommandtwoopt{\citeads}[3][][]{\href{http://adsabs.harvard.edu/abs/#3}%
    {\def\hyper@linkstart##1##2{}%
     \let\hyper@linkend\@empty\citealp[#1][#2]{#3}}}
  \newcommandtwoopt{\citepads}[3][][]{\href{http://adsabs.harvard.edu/abs/#3}%
    {\def\hyper@linkstart##1##2{}%
     \let\hyper@linkend\@empty\citep[#1][#2]{#3}}}
  \newcommandtwoopt{\citetads}[3][][]{\href{http://adsabs.harvard.edu/abs/#3}%
    {\def\hyper@linkstart##1##2{}%
     \let\hyper@linkend\@empty\citet[#1][#2]{#3}}}
  \newcommandtwoopt{\citeyearads}[3][][]%
    {\href{http://adsabs.harvard.edu/abs/#3}
    {\def\hyper@linkstart##1##2{}%
     \let\hyper@linkend\@empty\citeyear[#1][#2]{#3}}}
\makeatother

\begin{document}

    \title{Connectivity between the solar photosphere and chromosphere in a vortical structure}
    \subtitle{Observations of multi-phase, small-scale magnetic field amplification}
    \titlerunning{Connectivity between the solar photosphere and chromosphere in a vortical structure}

   \author{S.M. D\'iaz-Castillo
          \inst{1,2},
          C.E. Fischer\inst{3,7}
          \and
          R. Rezaei\inst{4}
          \and
          O. Steiner\inst{1,5}
          \and
          S. Berdyugina\inst{1,5,6}
          }

   \institute{Institut f\"ur Sonnenphysik (KIS),
              Georges-K\"ohler-Allee 401 A, Freiburg i.Br., Germany \\
              \email{smdiazcas@leibniz-kis.de}
        \and Faculty of Physics, University of Freiburg, Freiburg, Germany
        \and National Solar Observatory (NSO), Boulder, CO, USA
        \and Departments of Physics, Sharif University of Technology, Tehran, Iran
        \and Istituto Ricerche Solari Aldo e Cele Daccò (IRSOL), Locarno, Switzerland 
        \and Faculty of Informatics, Università della Svizzera italiana, Locarno, Switzerland
        \and European Space Agency (ESA), European Space Astronomy Centre (ESAC), Madrid, Spain
        }

   \date{Received XXX; accepted XXX}

 
  \abstract
   {High-resolution solar observations have revealed the existence of small-scale vortices, as seen in chromospheric intensity maps and velocity diagnostics. Frequently, these vortices have been observed near magnetic flux concentrations, indicating a link between swirls and the evolution of the small-scale magnetic fields. Vortices have also been studied with magneto-hydrodynamic (MHD) numerical simulations of the solar atmosphere, revealing their complexity, dynamics, and magnetic nature. In particular, it has been proposed that a rotating magnetic field structure driven by a photospheric vortex flow at its footprint produces the chromospheric swirling plasma motion.}
   {We present a complete and comprehensive description of the time evolution of a small-scale magnetic flux concentration interacting with the intergranular vortex flow and affected by processes of intensification and weakening of its magnetic field. In addition, we study the chromospheric dynamics associated with the interaction, including the analysis of a chromospheric swirl and an impulsive chromospheric jet.}
   {We studied observations taken with the CRisp Imaging SpectroPolarimeter (CRISP) instrument and the CHROMospheric Imaging Spectrometer (CHROMIS) at the Swedish Solar Telescope (SST) in April 2019. The data were recorded at quiet-Sun disc centre, consisting of full Stokes maps in the \ion{Fe}{i} line at $6173\,\AA$ and in the \ion{Ca}{ii} infrared triplet line at $8542\,\AA$, as well as spectroscopic maps in the lines of H$\alpha$ $6563\,\AA$ and \ion{Ca}{ii} K $3934\,\AA$. Utilising the multi-wavelength data and performing height-dependent Stokes inversion, based on methods of local correlation tracking and wavelet analysis, we studied several atmospheric properties during the event lifetime. This approach allowed us to interpret the spatial and temporal connectivity between the photosphere and the chromosphere.}
   {We identified the convective collapse process as the initial mechanism of magnetic field intensification, generating a re-bound flow moving upwards within the magnetic flux concentration. This disturbance eventually steepens into an acoustic shock wave that dissipates in the lower chromosphere, heating it locally. We observed prolonged magnetic field amplification when the vortex flow disappears during the propagation of the upward velocity disturbance. We propose that this type of magnetic field amplification could be attributed to changes in the local vorticity. Our analysis indicates the rotation of a magnetic structure that extends from the photosphere to the chromosphere, anchored to a photospheric magnetic flux concentration. It appears to be affected by a propagating shock wave and its subsequent dissipation process could be related to the release of the jet.}
   {}


   \maketitle
%

\section{Introduction}

Quiet-Sun regions harbour a pervasive and evolving small-scale magnetic field, organised across a range of different granular scales due to horizontal convective flows \citep[see e.g.][]{Solanki1993, DeWijn2009, BellotRubio2019}. At the supergranular scale, two magnetic field structures can be identified in photospheric magnetograms: 1) a kiloGauss field accumulation forming a network-like pattern, the so-called magnetic network \citep{Sheeley1967} and 2) the region between them containing innumerable weaker and smaller magnetic flux concentrations, the so-called magnetic internetwork \citep{BellotRubio2019}. In the internetwork, magnetic flux concentrations have typical sizes of less than $1^{\prime\prime}$. They are commonly accumulated in regions of downwardly directed plasma flow surrounding the granules, namely, within the intergranular space. These flux concentrations can be spatially associated with features of increased continuum brightness compared to their surroundings, the so-called magnetic bright points (MBPs) \citep{Solanki1993, SanchesAlmeida2004}. Advected by convective horizontal movements, these elements eventually interact with other magnetic fields and plasma, disappearing via the flux cancellation of opposite polarity pairs, or otherwise becoming stable magnetic flux concentrations via flux expulsion \citep{Parker1963, Weiss1963, GallowayandWeiss1981}. 

Network and internetwork magnetic structures have their counterpart in the chromosphere, and their dynamics highly influence the upper layers of the solar atmosphere. The gas pressure dominates the physical processes in the surface layers where the magnetic field is confined to individual and separated flux tubes. With the density decreasing with height, the magnetic flux tubes expand to the upper layers, filling the available space either merging or interconnecting with other accumulations of magnetic flux of equal or opposite polarities, respectively \citep{SolankiSteiner1990, Solanki1993}. The plasma regime change determines the coupling and connectivity between the photosphere and the chromosphere; therefore, the analysis of physical quantities based on layer diagnostics can lead us to understand the interrelation and causality of small-scale solar phenomena at different heights. For instance, transient events such as sudden amplification or disappearance of magnetic flux can produce energetic effects that can be observed in several layers of the solar atmosphere \citep[and references therein]{Chen2021}. 

Short-lived magnetic field intensification occurs in the internetwork magnetic flux, reaching kG field-strengths values significantly larger than the expected equipartition field strength \citep{Nagata2008}. The convective collapse, suggested initially by \cite{Parker1978}, is a proposed mechanism to reach such high field strengths from an initially weak magnetic field. In an incipient magnetic flux tube, the energy transport is inhibited leading to a thermally isolated tube. As a consequence of the super-adiabatic stratification of the surrounding medium, the down-flowing plasma accelerates and evacuates the tube, leading to its compression and thus its magnetic field intensification. This process was also identified using numerical simulations \citep{Hasan1985, Grossmann-Doerth1998, Takeuchi1999, Vogler2005}. In particular, \cite{Grossmann-Doerth1998} reported that the intensification is associated with a partial evacuation of the magnetic flux tube and an accelerating downflow, the effect previously described in \cite{Parker1978}. In their simulations, an initial magnetic field of around $400$~G leads to a fast downflow within magnetic flux concentration, causing a rebound effect. This produces a rising plasma flow that may trigger an upward-moving shock, which can then destroy the compressed flux tube.

From an observational point of view, based on the evidence given in \cite{BellotRubio2001} from ground-based spectropolarimetric observations, the convective collapse and the subsequent dispersal of the magnetic flux are due to the formation of an upward-moving front. An increase in the total circular polarisation measurements, together with a redshifted intensity profile in photospheric lines, was identified in the initial stage of the convective instability followed by a blue-shifted intensity profile and a sudden appearance of an extra displaced Stokes-$V$ component, producing a strong asymmetry in the Stokes-$V$ profile due to the upward-moving front. Using Hinode observations, \cite{Nagata2008} identified a small-scale magnetic element with an initial magnetic field strength of around 400 G with a transient intense downflow close to $6\unit{km\,s^{-1}}$, subsequently reaching a magnetic field strength of $2$~kG. Statistical studies were also performed demonstrating the ubiquitous occurrence of the convective collapse \citep{Fischer2009, Utz2014, Keys2020}. For instance, \cite{Fischer2009} studied 49 convective collapse events exhibiting a brightening in the continuum intensity, downward plasma flows in the photosphere and, in several cases, a brightening was seen in the \ion{Ca}{ii} K chromospheric line. A recent study of high-resolution spectropolarimetric observations of hundreds of MBPs including a comparison with radiative magneto-hydrodynamic (MHD) simulations revealed other mechanisms of sudden magnetic field amplification, closely related to convective motions \citep{Keys2020}. Using observational and numerical evidence, \cite{Keys2020} found that compression due to granular expansion and merging with surrounding magnetic accumulations can also produce a fast intensification of magnetic fields. They detected specific cases of magnetic field intensification at locations of enhanced vorticity, but only observed it in simulations. These findings suggest the importance of observational studies of magnetic field intensification within vortical flows.

Although diverse small-scale vortical motions in the photospheric intergranular regions have been frequently detected in solar observations \citep[e.g.][]{Brandt1988, Bonet2008, Bonet2010, Attie2009, Steiner2010, Vargas2011, Giagkiozis2018, Requerey2018}, it has been shown that they are not necessarily always co-located with magnetic field concentrations \citep{Giagkiozis2018}. However, it was found that they can influence the dynamics of neighbouring magnetic elements \citep{Bonet2008, Vargas2015} and control the plasma dynamics as they can channel mass, momentum, and energy into different solar atmospheric layers \citep{Wedemeyer2009, Morton2013, Tziotziou2018, Tziotziou2019, Liu2019, Tziotziou2020, Tziotziou2023, Jess2023}. At the network scale, \cite{Requerey2018} studied the relationship between a persistent photospheric vortex flow and the evolution of a magnetic network element using Hinode data. These authors showed that magnetic stability is closely related to the presence of vortex flows, a relationship that was theoretically predicted by \cite{Schuessler1984} and \cite{Buente1993}. Furthermore, \cite{Shetye2019} inspected small-scale chromospheric swirling patterns and their connection with the dynamics of underlying photospheric magnetic flux concentrations. They found that the swirl centre is co-spatial with single magnetic flux concentrations, suggesting that the vertical magnetic field of the concentration supports the swirl. Moreover, the magnetic nature of the vortical movements in the chromosphere was confirmed by later observations \citep{Murabito2020}.

Simulations have confirmed a large number density and several properties of vortical motions as well as their magnetic nature at intergranular scales \citep[e.g.][]{Moll2011, Shelyag2011, Silva2020, Silva2021, Battaglia2021, Tziotziou2023, CaniveteCuissa2024}. However, the study of their dynamics and multi-layer connectivity at the observational level has been challenging due to the difficulty of finding isolated cases and the diverse constraints in the available observational data. Fortunately, instruments such as the CRisp Imaging SpectroPolarimeter \citep[CRISP, ][]{Scharmer2008, Lofdahl2021} and the CHROMospheric Imaging Spectrometer \citep[CHROMIS, ][]{Scharmer2017, Lofdahl2021} attached to the Swedish Solar Telescope \citep[SST, ][]{Scharmer2003} are equipped to provide high-quality data products for such an analysis. 

In this work, we present a comprehensive description of the evolution of an isolated photospheric small-scale magnetic element undergoing sequential episodes of magnetic field intensification during its interaction with a vortex flow observed by CRISP and CHROMIS. The magnetic flux concentration evolves in connection with a chromospheric swirl, which features an impulsive plasma release simultaneously with one of the events of magnetic field intensification. In particular, we analyse the relationship of the studied case with a detected photospheric convective collapse process and its connectivity with a chromospheric jet propagating through a preexistent chromospheric swirl structure. Also, we have explored the heating signatures as well as evidence of high-frequency oscillations in the chromosphere.

\section{Data and methods}
\label{methods}

\begin{figure*}
\centering
\includegraphics[width=\textwidth]{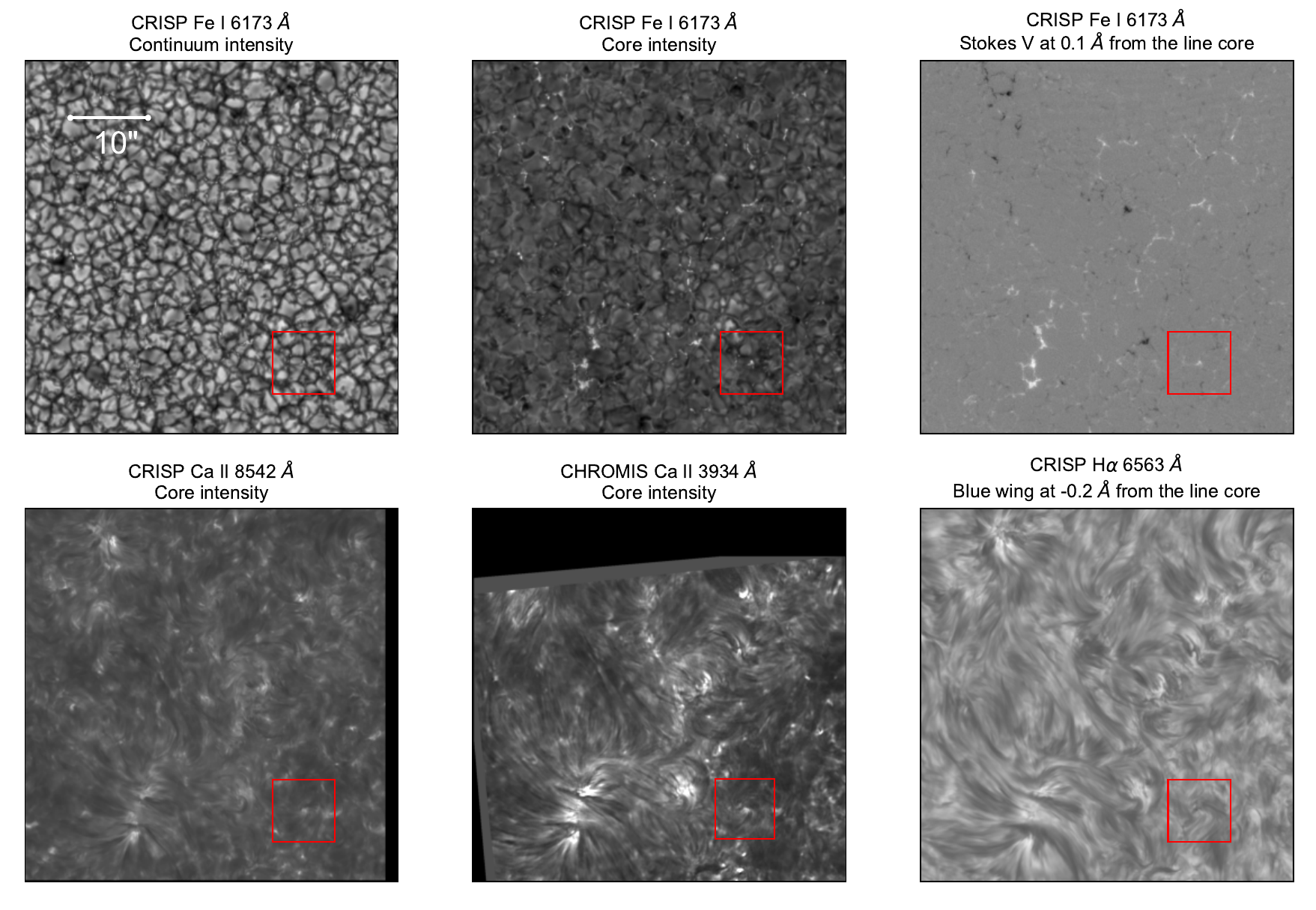}
\caption{Snapshots of the complete FOV ($50^{\prime\prime} \times 50^{\prime\prime}$) of CRISP and CHROMIS at a selected wavelength position in the sampled spectral lines: Top row from left to right: Continuum and core intensity of the spectral line \ion{Fe}{i} $6173.3\,\AA$, Stokes-$V$ circular polarisation in the red wing of the \ion{Fe}{i} $6173.3\,\AA$ line. Bottom row from left to right: Intensity of the core of Ca II at $8542.1\,\AA$ and $3933.7\,\AA$, and blue wing of H$\alpha$. The snapshots were recorded on April 24 2019 at 08:46:52UT when the event was initially detected. Red boxes mark a region of $8^{\prime\prime} \times 8^{\prime\prime}$ size, where the studied event is located.}
\label{Fig0}
\end{figure*}

The case under study was identified in simultaneous and cospatial full-Stokes observations of the \ion{Fe}{i} $6173.3\,\AA$ line (Landé-factor $g_{\rm eff}$ = 2.5) and the \ion{Ca}{ii} infrared triplet at $8542.1\,\AA$ (Landé-factor $g_{\rm eff}$ = 1.1) together with spectroscopy observation of the  H$\alpha$ line, all recorded with CRISP in the quite-Sun disc-centre. These observations were undertaken in April 2019 as part of a campaign run by the SOLARNET access programme. The mean seeing parameter $r_{0}$ was 22.7 cm during the 2 hours of the sequence. The data provided were corrected for dark current and flat-fielding using the SSTRED data pipeline \citep{Lofdahl2021}. The data were further demodulated and corrected for cross-talk (Stokes $I$ to $Q$, $U$, and $V$). A reconstruction process was performed using a multi-object multi-frame blind deconvolution (MOMFBD) algorithm, removing the small-scale atmospheric distortions to obtain high spatial resolution and precise alignment between the sequentially recorded images \citep{Lofdahl2002, VanNoort2005}.

The CRISP data set is composed of spatially aligned maps on a specific range of wavelength positions for each line. The \ion{Fe}{i} $6173.3\,\AA$ line is sampled in 15 wavelength points spread symmetrically around the line core in $0.035\,\AA$ steps, and the \ion{Ca}{ii} $8542.1\,\AA$ line is sampled in 21 wavelength points with a varying step size of $0.075\,\AA$ close to the core of the line, and up to $0.1\,\AA$ in the wings. The H$\alpha$ line is sampled in 25 wavelength points spread symmetrically around the line core in $0.1\,\AA$ steps. The complete field of view (FOV) is approximately $50^{\prime\prime} \times 50^{\prime\prime}$ with a spatial resolution of $0.12^{\prime\prime} \times 0.12^{\prime\prime}$ ($0.058^{\prime\prime}\text{pix}^{-1}$). The maps of the data set were recorded with a cadence of 28.2 seconds. The calculated photon noise in the polarimetric data (Stokes $Q$, $U$, and $V$) is approximately $\sigma_{p} = 2.3 \times 10^{-3}$ (in units of the spatially and temporally averaged continuum intensity, $I_c$).

The CRISP data set is complemented with simultaneous observations of the \ion{Ca}{ii} K line recorded with the CHROMIS instrument. The \ion{Ca}{ii} K line is sampled with 27 wavelength points around the line centre at $3933.7\,\AA$ with one continuum point at $3930\,\AA$. The wings of the line (blue and red) are sampled with three points at $\pm 0.92\,\AA$, $\pm 1.18\,\AA$, and $\pm 1.51\,\AA$ and the line core is sampled within the interval $-0.65\,\AA$ to $+0.65\,\AA$ with $0.065\,\AA$ spacing. Unlike the CRISP instrument, CHROMIS provides a cadence of 7.8 seconds, and a FOV covering our area of interest with a spatial resolution of $0.08^{\prime\prime} \times 0.08^{\prime\prime}$ ($0.039^{\prime\prime} \text{pix}^{-1}$). The calculated photon noise of the CHROMIS spectroscopic data is approximately $\sigma_{s} = 0.02$ (in units of the spatially and temporally averaged continuum intensity $I_c$). The details of the optical setup, along with the passbands of the different filters, are given in \cite{Lofdahl2021}.

The complete FOV was analysed using the Milne-Eddington (ME) inversion technique: Very Fast Inversion of the Stokes Vector (VFISV) \citep{Borrero2011} applied on the \ion{Fe}{i} $6173.3\,\AA$ line. We performed the wavelength calibration by comparing the mean spectral profile of the full quiet Sun region with the McMath-Pierce Fourier Transform Spectrometer atlas \citep[hereafter FTS atlas, ][]{Neckel1999}. Then, we applied VFISV to estimate the photospheric average parameters, such as the line-of-sight (LOS) velocity ($v_{\rm{LOS}}$) and magnetic field strength ($|B|$). The p-mode signals were removed in the LOS velocity using a pass-band filter. Using the magnetic field strength maps, we tracked every ephemeral magnetic field concentration above $1$~kG in the full sequence with the Yet Another Feature Tracking Algorithm (YAFTA) \citep{Welsch2003, DeForest2007}. From hundreds of magnetic flux concentrations, most of those localised in network patches, we identified our studied event. As a context image, Fig.\,\ref{Fig0} depicts the complete FOV at different wavelength positions of the spectral lines sampled by CRISP and CHROMIS. The snapshots presented were recorded on 24.04.19 at 08:46:52UT when the event was initially detected. Red boxes mark a region of $8^{\prime\prime} \times 8^{\prime\prime}$ size, where the studied event is located.

\begin{table}
      \caption[]{Definition of the characteristic parameters and spectral band samples of the \ion{Ca}{ii} K line core.}
         \label{t1}
         \[\begin{array}{p{0.3\linewidth}ll}
            \hline
            \noalign{\smallskip}
            \text{Band sample} & \text{Central wavelength[\AA]} & \text{Width[\AA]} \\
            \noalign{\smallskip}
            \hline
            \noalign{\smallskip}
            K-\text{index}     & 3933.7 & \pm 0.52    \\
            \text{$K_{2V}$}      & 3933.5 & \pm 0.065   \\
            \text{$K_{2R}$}      & 3933.9 & \pm 0.065   \\
            \text{$K_3$}       & 3933.7 & \pm 0.065   \\
            \noalign{\smallskip}
            \hline
            \hline
            \noalign{\smallskip}
            \text{Parameter} & \text{Definition} & \\
            \noalign{\smallskip}
            \hline
            \noalign{\smallskip}
            \text{Emission strength}   &   K_{2V}/K_{3} &   \\
            \text{\text{$K_{2}$}-asymmetry}     &   (K_{2V}-K_{2R})/(K_{2V}+K_{2R}) &   \\
            \noalign{\smallskip}
            \hline
         \end{array}\]
    \end{table}

We applied several methods for the analysis of this special small-scale magnetic flux concentration. First, we used the VFISV results to describe the average physical properties in the photosphere. We also explored the magnetic field characteristics of the magnetic flux concentration around the formation height of the \ion{Fe}{i} $6173.3\,\AA$ line using the spatial distribution of the total circular polarisation (TCP), the total linear polarisation (TLP), and the net circular polarisation (NCP) defined as follows:
\begin{align}
\text{TCP} =& \frac{1}{N} \int_{\lambda_{\rm{i}}}^{\lambda_{\rm{f}}} \frac{\sqrt{V^2}}{I_{\rm{c}}} d\lambda,\\
\text{TLP} =& \frac{1}{N} \int_{\lambda_{\rm{i}}}^{\lambda_{\rm{f}}} \frac{\sqrt{Q^2+U^2}}{I_{\rm{c}}} d\lambda,\\
\text{NCP} =& \frac{V_{\rm{b}} - V_{\rm{r}}}{V_{\rm{b}} + V_{\rm{r}}},
\end{align}
where $N$ is the number of total spectral points in the profile, $V_{\rm{b}} = {N_{\rm{b}}}^{-1} \lvert \int_{\lambda_{\rm{i}}}^{\lambda_{\rm{zc}}} V d\lambda \rvert /I_{\rm{c}}$ and $ V_{\rm{r}} = {N_{\rm{r}}}^{-1} \lvert \int_{\lambda_{\rm{zc}}}^{\lambda_{\rm{f}}} V d\lambda \rvert /I_{\rm{c}}$ are the normalised circular polarisation integrated over the blue (from $\lambda_{\rm{i}}$ to ${\lambda_{\rm{zc}}}$) and red (from $\lambda_{\rm{zc}}$ to ${\lambda_{\rm{f}}}$) wavelength range of the profile respectively, taking as reference the Stokes-$V$ zero-crossing wavelength position (${\lambda_{\rm{zc}}}$) and considering $N_{\rm{b}}$ and $N_{\rm{r}}$ the number of spectral points in the blue range and in the red range respectively. The NCP analysis provides evidence of gradients of the physical parameters within the magnetic flux concentration \citep{Landolfi1996}. 

We complement this analysis by performing height-dependent inversion using the Stokes Inversion based on Response functions (SIR) algorithm \citep{RuizCobo1992} applied on the \ion{Fe}{i} $6173.3\,\AA$ line to infer the atmospheric physical parameters in different photospheric heights. For this spectral data, all velocities and Stokes-$V$ zero-crossing shifts presented refer to the core wavelength position of the mean Stokes-$I$ profile calculated for the complete FOV and time sequence. The SIR algorithm setup, inputs, and approach are described at the beginning of Sect.\,\ref{SIR}.

\label{Photosphere}
\begin{figure}
   \centering
   \includegraphics[width=\columnwidth]{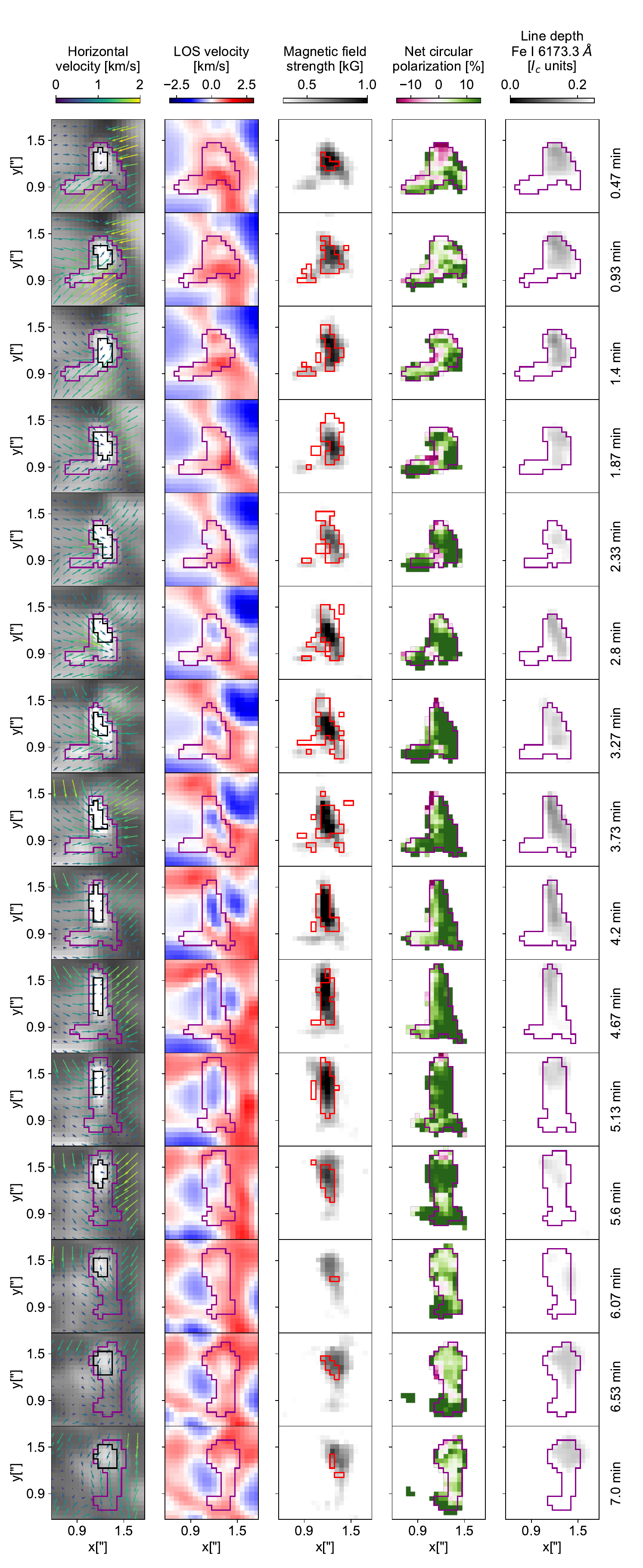}
   \captionsetup{textfont=small}
   \caption{Evolution (top to bottom) of the studied magnetic flux concentration in the photosphere (see the time stamp in the right margin). First column: Continuum intensity including the horizontal component of the velocity field, where each arrow represents $3\times3$ pixels. The length and the colour of the arrows are proportional to the magnitude of the velocity. The purple contour highlights the full area of the MBP. The black contour refers to the area of the MBP core. Second and third column: LOS velocity and magnetic field strength from the VFISV inversion. Red contours in the third column correspond to TLP above $1.5\sigma_p$. Fourth column: NCP percentage. Fifth column: Line depth of the \ion{Fe}{i} line in units of the continuum intensity.}
   \label{Fig1}%
\end{figure}

We estimated the horizontal velocity distribution based on the continuum images using the DeepVel model \citep{AsensioRamos2017}. This neural network model was trained using synthetic continuum images from magneto-convection simulations and tested with continuum intensity images from the Sunrise/IMAX data (spatial resolution $48\unit{km\,\text{pix}^{-1}}$) \citep{Martines2011}, which is slightly larger than the SST pixel size. However, the authors reported that this model can generalise correctly, independent of structure sizes. A cadence of 30 seconds was used for training and testing the model, 1.8 seconds less than the CRISP cadence, which may the magnitude of the horizontal velocities to become slightly underestimated.

For the chromospheric lines, we inspected the spatial distribution of the intensity of the \ion{Ca}{ii} $8542.1\,\AA$ line core and we characterised the plasma flows in the lower and middle chromosphere, using bisector analysis of the intensity profiles. We compared these results with wavelength-integrated intensity maps calculated over specific spectral band samples of the \ion{Ca}{ii} K line core, using a similar approach as presented by \cite{Rezaei2007_2}. In particular, we used the K-index parameter defined as the intensity average in a $1\AA$ spectral band centred on the line core, the emission-strength parameter, and the intensity excess of the $K_{\rm{2V}}$ and $K_{\rm{2R}}$ bands as proxies of chromospheric heating. We also inferred the directional sense of the plasma flow in the upper chromosphere using the asymmetry between the $K_{\rm{2V}}$ and $K_{\rm{2R}}$ bands. Table \ref{t1} specifies the definition of these spectral bands and the computed parameters.

In addition, we inspected the existence of high-frequency oscillatory signatures in the intensity of defined spectral bands of the \ion{Ca}{ii} K line core, leveraging on the high cadence of the observations. Based on wavelet analysis of the time series, we explored the power contribution for signals above 5.5 mHz, higher than the cutoff frequency in the chromosphere \citep{Carlsson1997}. The results of this study are presented in the Appendix \ref{wavelets}.

As a tool for characterising similar profile behaviour in the maps, we used the k-means classification method \citep{Jin2010} for the profile classification for Stokes $I$ and $V$ of the calcium lines. 

Finally, we inspected the H$\alpha$ line to obtain the overall flow structure of the middle and upper chromosphere using bisectors and centre-of-gravity (COG) velocity maps. We compute the maps of full-width-half-maximum (FWHM) and equivalent width (EW) of the H$\alpha$ line to infer heating signatures in the chromosphere. The FWHM is calculated for each profile of the H$\alpha$ line as the distance between the wavelengths on either side of the line where the intensity reaches the half-maximum intensity, $I_{hm}$. This intensity is defined as $I_{hm}=(I_{min}+I_{c})/2$, where $I_{min}$ is the minimum intensity of the line and $I_{c}$ is the average between the intensities measured at $6561.6 \AA$ and at $6564.0 \AA$, corresponding to the first and last wavelength positions sampled. The EW is calculated for each profile as follows: 
\begin{equation}
    EW = \Delta\lambda \sum_{\lambda_i}(I_{max} - I_{\lambda_i}),
\end{equation}
where $\Delta\lambda$ is the wavelength step, $I_{max}$ is the maximum intensity in the line, and $I_{\lambda_i}$ is the intensity as function of wavelength position ($\lambda_i$). The intensity is normalised to the continuum.

\section{Results}

\subsection{Photospheric synopsis}
\label{ME}

In this subsection, we present an overview of the photospheric appearance of the MBP under investigation in the form of a time series of maps of various physical quantities, which are derived from the continuum intensity and the photospheric spectral line \ion{Fe}{i} $6173.3\,\AA$.

The magnetic flux concentration is initially identified with the YAFTA feature tracking algorithm lasting approximately 8 minutes before it dissolves. The initial flux expulsion phase or previous convective collapse process is not studied because previous frames have a low quality (low $r_0$ values) rendering the analysis questionable. Figure\,\ref{Fig1} shows (from top to bottom) the high-quality temporal evolution of the magnetic flux concentration together with several inferred quantities. 

\begin{figure}
   \centering
   \includegraphics[width=9cm]{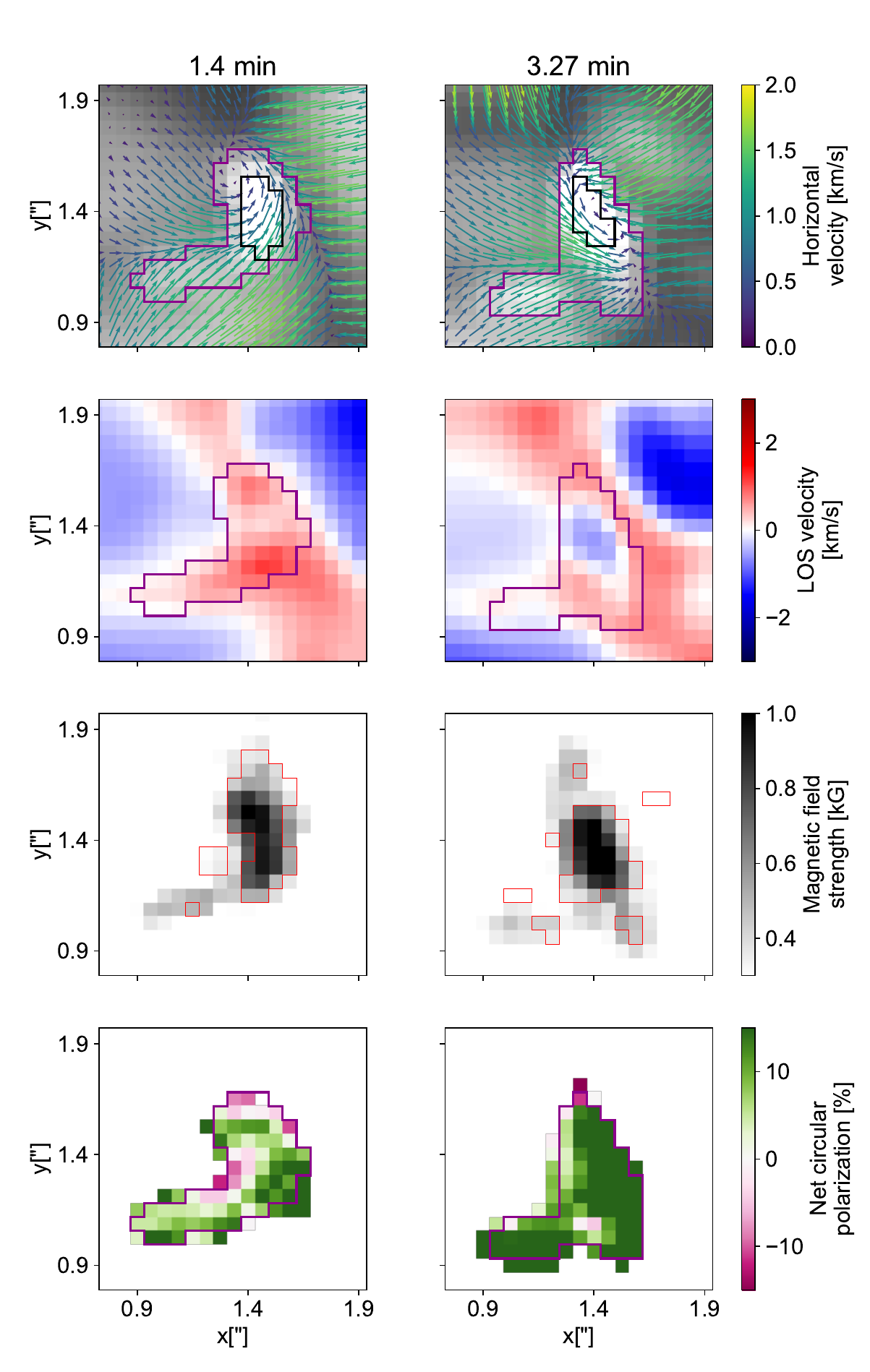}
   \captionsetup{textfont=small}
   \caption{Zoomed-in maps of the 1.4 and 3.27 min time steps from Fig.\ref{Fig1}. First row: Continuum intensity including the horizontal component of the velocity field, where arrows are plotted pixel by pixel. The length and the colour of the arrows are proportional to the magnitude of the velocity. Second row: LOS velocity. Third row: Magnetic field strength including contours of linear polarisation. Fourth row: Distribution of the percentage of the NCP. Contours have the same meaning as in Fig.\,\ref{Fig1}.}
   \label{Fig2}%
\end{figure}

The maps (in the first column) of Fig.\,\ref{Fig1} show the continuum intensity as background and the inferred horizontal velocity given by DeepVel as coloured arrows. The magnetic flux concentration reveals itself with an enhanced continuum intensity in comparison to the intergranular intensity classifying it as an MBP. The purple countors delimit the MBP area given by a threshold of $400$~G, while the black contours bound the MBP core conformed by the brightest $7\%$ pixels within the MBP area in each frame. The purple contours are drawn on several of the other columns to indicate the position of the MBP. Based on the purple contours, the MBP area exhibits the shape of a logarithmic spiral during the initial two minutes of the sequence. Afterwards, it deforms, apparently affected by granular buffeting. In some time steps before $t = 3.27\,\text{min}$, the horizontal velocity pattern shows a counterclockwise intergranular vortex flow localised within the MBP area and coherent with its spiral shape. Figure \ref{Fig2} depicts a zoomed view for two specific time steps when the counterclockwise intergranular vortex flow is co-spatial to the MBP area. 

The time series of the LOS velocity from the VFISV inversion is shown in the second column of Fig.\,\ref{Fig1}. The usual LOS velocity of the granulation is identifiable in the surroundings of the magnetic flux concentration, meaning positive LOS velocities (red) in the intergranular region where the plasma is moving downwards (line redshift), and negative LOS velocities (blue) within granules where the plasma is moving upwards (line blueshift). The magnetic flux concentration is located within the intergranular space, thus its area is dominated by downflows in the initial two minutes of the sequence, reaching speeds around $1\unit{km\,\text{s}^{-1}}$. However, an up-flowing component appears in the MBP centre at $t=2.8\,\text{min}$ lasting three minutes with a maximum average speed of $-2\unit{km\,\text{s}^{-1}}$ found at $t=4.2\,\text{min}$. The up-flowing component evolves in agreement with the deformation of the MBP until $t=5.6\,\text{min}$ and then disappears giving way again to a downflow. 

The time series of the magnetic field strength from the VFISV inversion is shown in the third column of Fig.\,\ref{Fig1} together with red contours marking signals of TLP above $1.5\sigma_p$, inferred with the polarimetric measurements of the \ion{Fe}{i} line. During the first and second minute of evolution, when the plasma moves downwards in the MBP area, the MBP core is characterised by magnetic field strengths around $1\,$kG, while in its full area, the magnetic field strengths are below $1\,$kG on average. From $t=2.8\,\text{min}$ onwards, lasting for three minutes, a prolonged magnetic intensification occurs simultaneously with the up-flowing component in the MBP area, reaching a maximum value of $1.6\,$kG with a spatial average of around $1\,$kG. Finally, when the downward flow sets in again after $t=6.07\,\text{min}$, the magnetic field strength drops until reaching an average value of $0.7\,$kG calculated over the MBP area. A marginal but spatially coherent region of TLP signals surrounds the magnetic flux concentration during several frames before $t = 5.13\,\text{min}$. Since it occurs at the periphery of the magnetic flux concentration it hints at a coherent transversal magnetic field component as one would expect from a twisted magnetic flux tube. However, given the noisy behaviour of the Stokes $Q$ and $U$ signals, and together with the limited spectral resolution given by Fabry-Perót spectrographs such as CRISP, it is not possible to rigorously study the transverse component of the magnetic field. 

We infer the spatial distribution of the degree of asymmetry of the Stokes-$V$ profile based on the NCP. The fourth column of Fig.\,\ref{Fig1} shows the distribution of the NCP percentage calculated in pixels with TCP signal greater than three times its variance, most of them localised inside the MBP area. At the beginning of the sequence, we identify positive and negative NCP components within the MBP area characterised as follows: At $t =0.47\,\text{min}$, a patch of negative NCP is located near the upper boundary of the MBP area, where the downflow is weak. After a minute at $t = 1.4\,\text{min}$, a patch of negative NCP appears in the central-left part of the MBP area, largely surrounded by pixels of positive NCP, resembling the case observed by \cite{Rezaei2007}. This behaviour was shown to be due to a canopy-like boundary layer, given by the expansion of the magnetic flux structure with height. The zoomed-in view of this time step is presented in the first column of Fig.\,\ref{Fig2}, showing that this boundary layer of positive NCP follows the counterclockwise intergranular vortex flow. Starting from $t = 2.8\,\text{min}$, a dominant positive NCP component above $10\%$ covers a significant portion of the MBP area, coinciding with the presence of the up-flowing component. Based on \cite{BellotRubio1997}, positive NCP values are compatible with sharp LOS velocity gradients arising from a flow disturbance propagating upwards. The low sensitivity of the line to the magnetic field compared with spectral lines in the infrared, and the limited spectral resolution do not allow us to identify the fine details of the Stokes-$V$ profiles as described by \cite{BellotRubio2001}. Thus, we can not resolve multiple lobes of the Stokes-$V$ profiles that may be associated with an upwardly moving front.

The last column of Fig.\,\ref{Fig1} shows the maps of the relative line depth of the \ion{Fe}{i} line. The area of the magnetic element exhibits an overall enhanced line-core intensity, however, in some particular time steps ($t = 1.4, 3.73, 4.2 \,\text{min}$) the absorption line is relatively flat (reaching a line depth of $0.12I_c$) for certain resolution elements inside the MBP area. The observation of the line weakening \citep{Sheeley1967, Beckers1968} in this magnetic sensitive line indicates opacity changes due to either magnetic field intensification with associated Zeeman broadening or localised heating in the upper photosphere.

To summarise, we observe the evolution of a MBP, initially embedded in a downwardly directed flow of counterclockwise vortical motion and a magnetic field strength of up to 1 kG, followed by an upwards flow during which the magnetic field strengthens up to 1.6 kG, and a final relaxing downflow phase in which the field weakens again.

\subsection{MBP fine-structure and magnetic field amplification}
\label{SIR}

\begin{table}
      \caption[]{SIR inversion parameters}
         \label{t2}
         \[\begin{array}{p{0.25\linewidth}lll}
            \hline
            \noalign{\smallskip}
                  &  \text{Cycle 1} & \text{Cycle 2} & \text{Cycle 3} \\
            \noalign{\smallskip}
            \hline
            \noalign{\smallskip}
            [I,Q,U,V] weights & [5,1,1,50]^* & [5,1,1,50]^* & [5,1,1,50]^*     \\
            T nodes                   & 3 & 5 & 7     \\
            |B| nodes                 & 1 & 2 & 2     \\            
            \text{$V_{LOS}$ nodes}           & 1 & 2 & 3     \\             
            \noalign{\smallskip}
            \hline
         \end{array}\]
    \footnotesize{$^*$ For cases of $Q$ and $U$ profiles with signals greater than $3\sigma_{p}$, we modify the weights to 5 for $Q$ and $U$.}
   \end{table}

\begin{figure*}
\centering
\includegraphics[width=\textwidth]{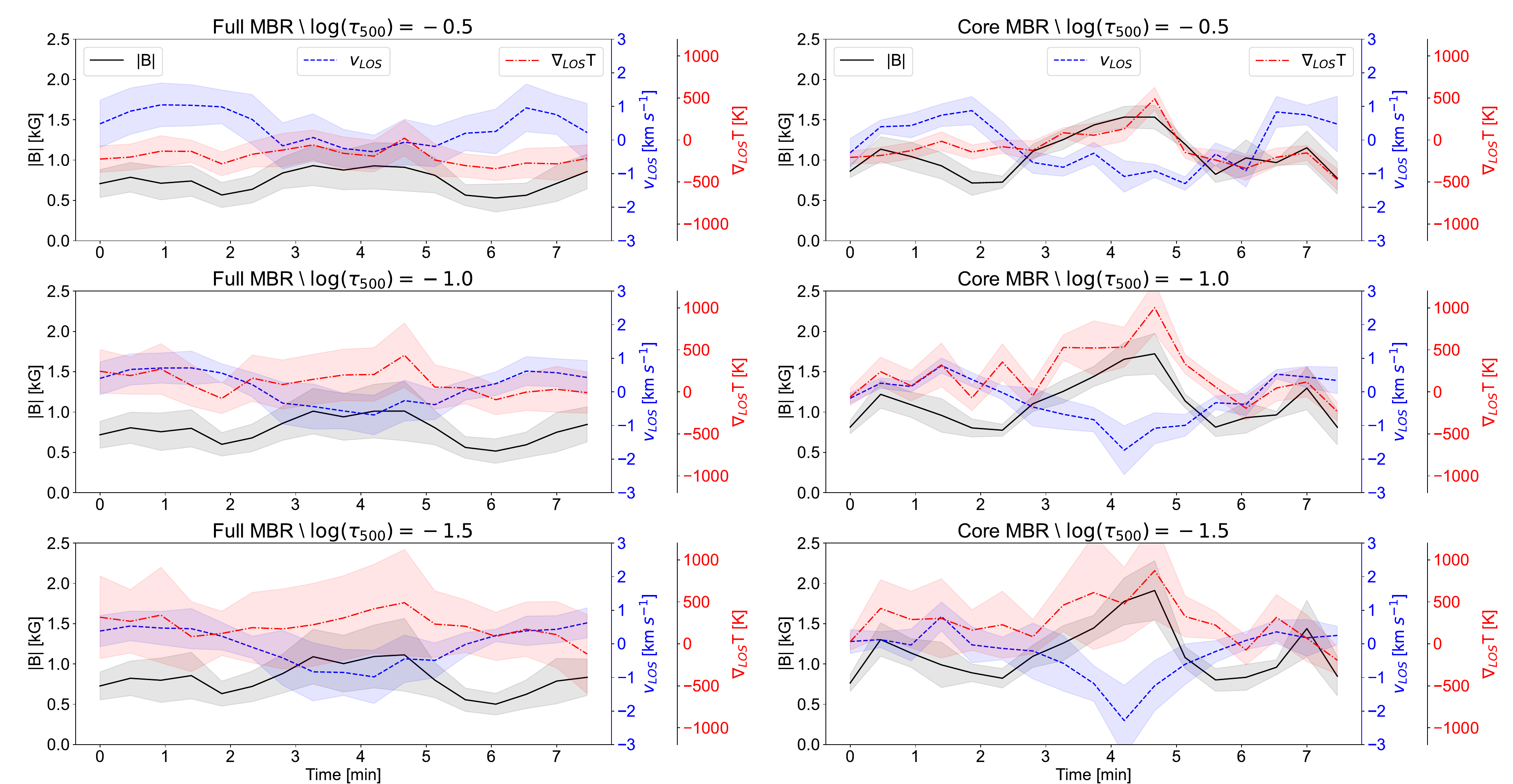}
\caption{Temporal evolution of the magnetic field strength $|B|$ (solid black lines), LOS velocity $v_{LOS}$ (dashed blue lines) and temperature LOS changes $\nabla_{LOS} T$ (dot-dashed red lines) at three different $\log \tau_{500}$ values estimated from SIR inversions of the \ion{Fe}{i} 617.3 nm profiles. First column: Results from the spatial average of the full area of the MBP (region marked by the purple contour in the first column of Fig.\,\ref{Fig1}). Second column: Results from the spatial average of the MBP core (region marked by the black contour in the first column of Fig.\,\ref{Fig1}). The shaded areas represent the standard deviation as a statistical error for each curve. The legends shown in the first row of plots apply to all plots.}
\label{Fig3}
\end{figure*}

To test potential heating processes at different photospheric heights and their relation to the magnetic field intensification, we study in this subsection the evolution and fine structure of the MBP, based on the height-dependent SIR inversion of the photospheric spectral line \ion{Fe}{i} $6173.3\,\AA$.

We selected the total area of the MBP corresponding to the region bounded by the purple contours in Fig.\,\ref{Fig1}. We performed the inversion per individual pixel within the area for all frames in the time sequence. Each inversion ran three cycles using the parameters given in Table \ref{t2}. As algorithm inputs, we took the CRISP spectral PSF and a local stray light profile corresponding to an average intensity profile of an external quiet region. As the initial atmosphere, we used the Harvard Smithsonian Reference Atmosphere model \citep[HSRA,][]{Gingerich1971} defined in a $\log \tau_{500}$ range spanning from 1.0 to -5.0 with a 0.1 sampling step. We left the macroturbulence, microturbulence, inclination, and azimuth of the magnetic field vector as free parameters, and constrained the magnetic filling factor to be one. The final inversions were selected based on the resulting $\chi^2$-merit function parameter. The analysis of these inversions was carried out carefully as we are aware of the limited spectral resolution and the high noise level of the profiles, which constrains the validity of the height-dependency of the inversions. Regarding the latter, the number of nodes in the magnetic field strength and the LOS velocity (see Table \ref{t2}) produces constant and linearly stratified atmospheres for the two first cycles respectively, with an extra node correction for the LOS velocity in the last cycle. With this approach, we avoided adding spurious results associated with the interpolation algorithm. The \ion{Fe}{I} photospheric line has the highest sensitivity to the atmospheric parameters around $\log \tau_{500} = -1.0$ based on a response function analysis \citep{Quintero2021}. Thus we expected a reliable result close to $\log \tau_{500} = -1.0$ and high uncertainties far from this optical depth. 

\begin{figure*}
\centering
\includegraphics[width=0.95\textwidth]{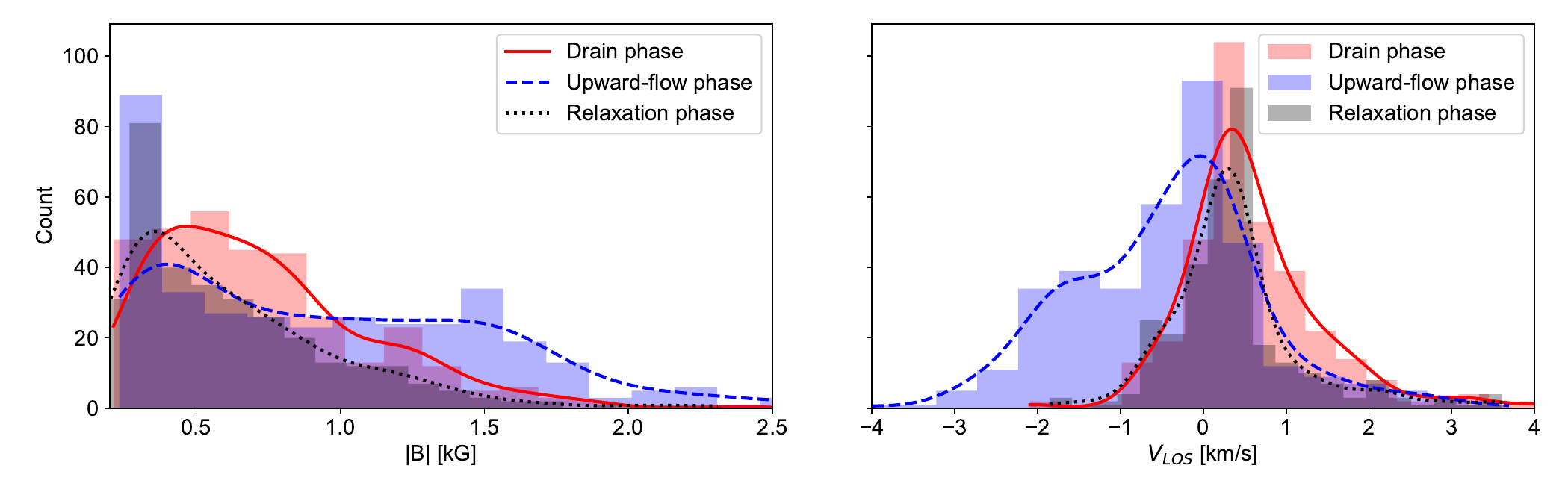}
\caption{Probability density function (histogram and its respective curve) for the magnetic field strength (left plots) and LOS velocity (right plots) over the area of the magnetic flux concentration (region marked by the purple contour in the first column of Fig.\,\ref{Fig1}) for the three defined evolution phases: the drain phase (between $t = 0.0\,\text{min}$ to $t = 2.3\,\text{min}$), the upward-flow phase (between $t = 2.8\,\text{min}$ to $t = 5.13\,\text{min}$), and the relaxation phase (between $t = 5.6\,\text{min}$ to $t = 7.0\,\text{min}$). The information included in the labels applies to both panels.}
\label{Fig4}
\end{figure*}

We made use of the spatial and temporal resolution of our data to study the time/height evolution of the fitted atmospheres returned by the SIR inversion. In particular, we are interested in the spatial-averaged behaviour of the selected physical parameters within the magnetic flux concentration. Using the best fit of the atmospheric models for every pixel, we calculated the spatial average of the physical parameters over two defined spatial domains of the magnetic flux concentration for each time step. The first spatial domain corresponds to the total area of the MBP bounded by the purple contours in Fig.\,\ref{Fig1} to skim its overall physical properties. The second spatial domain corresponds to the MBP core bounded by the black contour in the first column of Fig.\,\ref{Fig1}. We studied the evolution of the magnetic field strength ($|B|$), the LOS velocity ($v_{LOS}$), and the temperature gradient as a function of height, averaged over the above-mentioned spatial domains (see Fig.\,\ref{Fig3}). The temperature gradient in height is defined as the difference in temperature between specific $\log \tau_{500}$ values, that is, $\nabla_{LOS} T = T(\log \tau_{500} = \alpha - 0.5) - T(\log \tau_{500} = \alpha)$, for $\alpha$ values of $-0.5$, $-1.0$ and $-1.5$.

The first column of Fig.\,\ref{Fig3} shows the time evolution of the considered physical parameters for the case of the first spatial domain, namely, the total area of the MBP. In this case, the area is relatively big with a lateral extension of around $0.7^{\prime\prime}$ composed of i) a central bright region that seems to accrete magnetic flux from the surroundings and ii) a bright spiral arm in the lower left area, which later straightens after $t = 4.2\,\text{min}$. For the three $\log \tau_{500}$ values selected, the magnetic field strength and LOS velocities (black solid line and blue dashed line in Fig.\,\ref{Fig3}, respectively) have an anti-correlated behaviour on average. The inversion results show no significant changes in the average magnetic field strength and the average temperature gradient between the three selected $\log \tau_{500}$ values, meaning that these parameters do not change much as a function of height. The average LOS velocity indicates a downward flow at the beginning of the sequence that decreases with height. 

The second column of Fig.\,\ref{Fig3} shows the results for the MBP core. With a lateral extension of $0.2^{\prime\prime}$, this region contains the strongest magnetic field. The evolution of the average magnetic field strength exhibits two peaks where values go above $1$ kG in two short periods: 1) 56 seconds from $t = 0.47\,\text{min}$ to $t = 1.40\,\text{min}$ and 2) 28 seconds around $t = 7.0\,\text{min}$, and one prolonged intensification of 140 seconds, starting at $t = 2.8\,\text{min}$ and ending at $t = 5.13\,\text{min}$ with the maximum at $t = 4.8\,\text{min}$. Thus, the core of the magnetic flux concentration undergoes repeated magnetic field intensification and weakening. During the two short periods of magnetic field intensification, the MBP is dominated by downflows before the detection, in agreement with the amplification scenario due to convective collapse \citep{BellotRubio2001}. However, during the prolonged intensification, we observe the following occurrences: 1) the vortex flow is no longer observable, its last detection being at $t = 3.27\,\text{min}$, 2) the highest value of the upward velocity of the plasma is detected at $t = 4.2\,\text{min}$, and 3) a large change of temperature of around $800$ K from $\log \tau_{500} = -1.0$ to $\log \tau_{500} = -1.5$ is measured at $t = 4.8\,\text{min}$, hinting at a possible heat deposition at the line core formation height or above.

The different behaviour of the physical parameters within the two analyzed spatial domains indicates the presence of distinct substructures within the magnetic flux concentration, which can be further studied by considering distinct time windows in the sequence. We identify three specific time phases in the photosphere: 
\begin{enumerate}
    \item an initial phase of a down-flowing vortical drain between $t = 0.0\,\text{min}$ and $t = 2.3\,\text{min}$;
    \item an upward-flow phase between $t = 2.8\,\text{min}$ and $t = 5.13\,\text{min}$;
    \item a relaxation phase from $t = 5.6\,\text{min}$ to $t = 7.0\,\text{min}$.
\end{enumerate}
Figure\,\ref{Fig4} shows the histogram of the magnetic field strength and the LOS velocity for all pixels in the total area of the MBP at $\log \tau_{500} = -1.0$ for each of the three phases. 

During the vortical drain phase (represented by the red bars and solid red curve in Fig.\,\ref{Fig4}), $80\%$ of the total area of the MBP is dominated by plasma moving downwards, with a median velocity of $0.42\,\unit{km\, s^{-1}}$. Compared with Fig.\,\ref{Fig1}, downflows in this phase occur mostly near the borders of the MBP area. The distribution of magnetic field strength during the drain phase exhibits a dominant component, where in $80\%$ of the pixels fall, between $250\,$G and $1\,$kG, with a median value of $650\,$G. Additionally, there is a minor strong field component consisting of pixels with values above $1\,$kG. Examining Fig.\,\ref{Fig1}, the spatial distribution of the LOS velocity does not show a clear correlation with the spatial distribution of magnetic field strength.

The upward-flow phase (blue bars and dashed curve in Fig.\,\ref{Fig4}) is characterised by two Gaussian components in the LOS velocity distribution, one centred on $0.0\unit{km\,\text{s}^{-1}}$ with a dispersion of around $1.0\unit{km\,\text{s}^{-1}}$, and the second centred on $-1.5\unit{km\,\text{s}^{-1}}$ with a dispersion of around $1.5\unit{km\,\text{s}^{-1}}$. Two components are also present in the magnetic field strength distribution but with a flattened shape, one weak component (below $1\,$kG), and the other intense component (above $1\,$kG). For this case, the intense component is spatially correlated with the second LOS velocity component and with the vortex pattern, namely, the fast upward flow is co-spatial and co-temporal with the intensified component of the magnetic field and with the vortex pattern as can be seen from Figs.\,\ref{Fig1}-\ref{Fig2}. 

In the relaxation phase (grey bars and dotted curve in Fig. \ref{Fig4}), the distribution of the LOS velocities exhibits a shape similar to that of the vortical drain phase, with a median value of $0.32\unit{km\,\text{s}^{-1}}$, but is less asymmetric. Additionally, the distribution of the magnetic field strength during this phase shows a dominant component below $500\,$G, indicating a weakening of the magnetic flux concentration.

In summary, the SIR inversion of the spectral line \ion{Fe}{i} $6173.3\,\AA$ confirms and refines the photospheric synopsis of \ref{ME}: the core of the MBP undergoes repeated magnetic field intensification and weakening, where the middle, principal intensification occurs in the upward-flow phase, in particular in the core of the MBP with speeds that increase with height.

\begin{figure*}
\centering
\includegraphics[width=\textwidth]{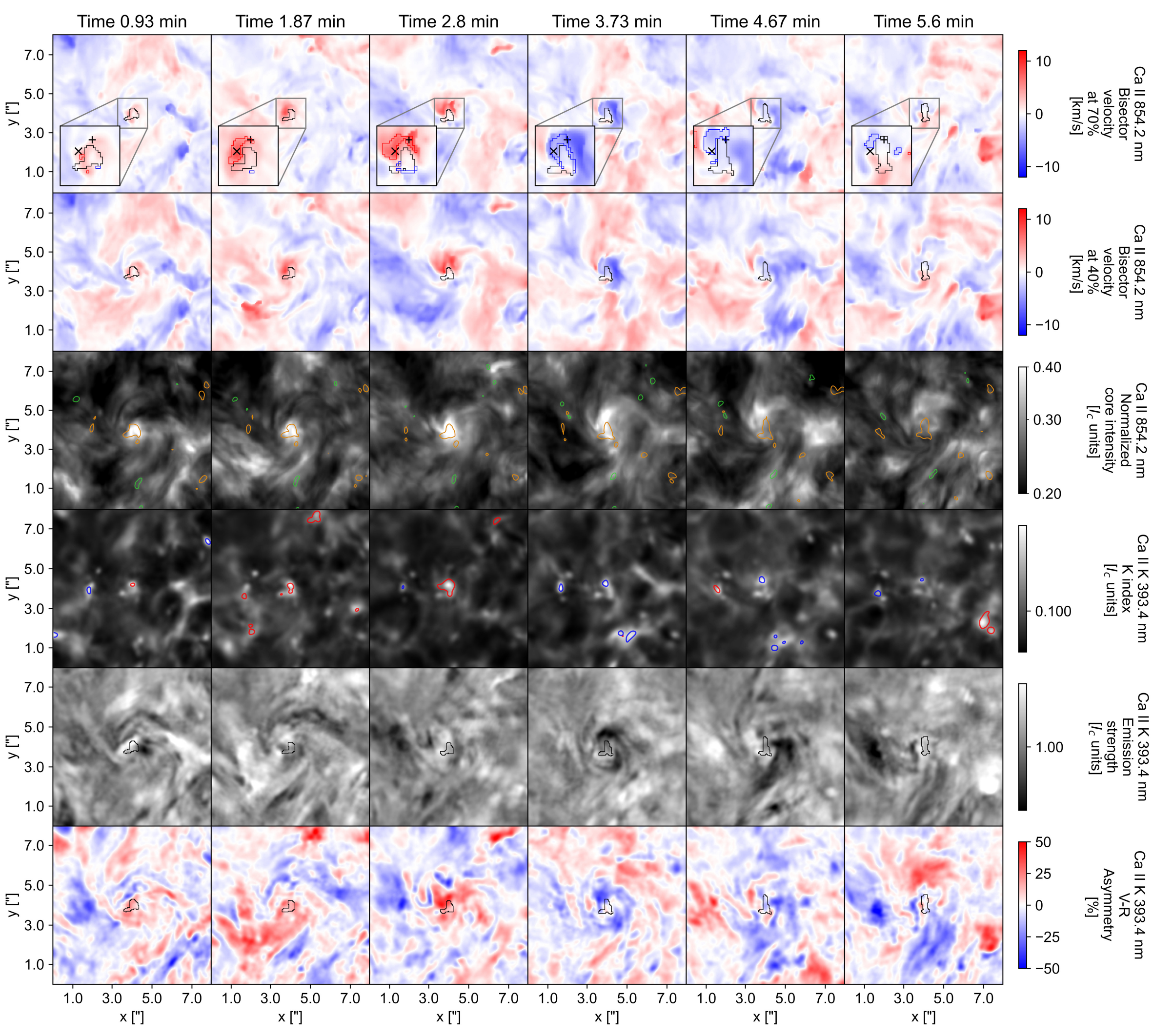}
\caption
{Time series of maps with a FOV of $8^{\prime\prime} \times 8^{\prime\prime}$ centred on the magnetic flux concentration showing several calculated quantities from the \ion{Ca}{II} spectra. First row: Bisector velocity at 70\% of the line depth of the \ion{Ca}{II} at $8542.1\,\AA$ line core. The maps include zoomed views ($1.5^{\prime\prime} \times 1.5^{\prime\prime}$) of the magnetic flux concentration where two points of interest (POIs) are indicated (POI\,1 black $\times$-sign and POI\,2 black $+$-sign). Red contours (drain phase) and blue contours (upward-flow phase) mark the areas of similar profiles of the POI\,1 shown in Fig.\,\ref{Fig6}. Second row: Bisector velocity at 40\% of the line depth of the \ion{Ca}{II} at $8542.1\,\AA$ line core. Third row: Normalised intensity of the core of the \ion{Ca}{II} line at $8542.1\,\AA$ in a logarithmic scale. Contours mark regions of intense (>2.5$\sigma_{p}$) TCP of the photospheric \ion{Fe}{i} line at $6173.3\,\AA$ (orange contours correspond to negative magnetic flux concentrations and light green contours correspond to positive magnetic flux concentrations). Fourth row: K-index defined in Table \ref{t1} in logarithmic scale including contours of intensity excess (>0.2$I_c$) of the $K_{2V}$ band (red) and the intensity excess of the $K_{2R}$ band (blue). Fifth row: Emission strength defined in Table \ref{t1} in logarithmic scale. Sixth row: $K_{2}$ asymmetry defined in Table \ref{t1} in percentages. Several maps include a central black contour to show the location, scale, and evolution of the MBP area. An animation is available online.
}
\label{Fig5}
\end{figure*}

\subsection{Chromospheric swirl}
\label{swirl}

\begin{figure*}
\centering
\includegraphics[width=0.49\textwidth]{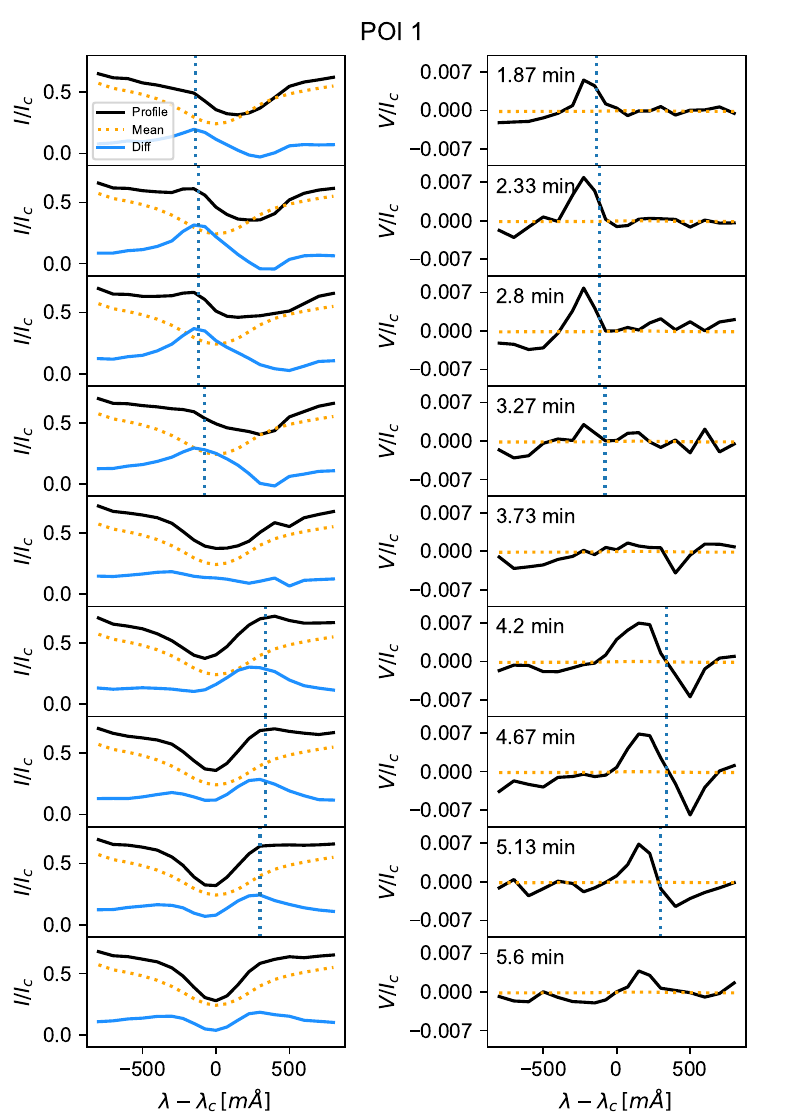}
\includegraphics[width=0.49\textwidth]{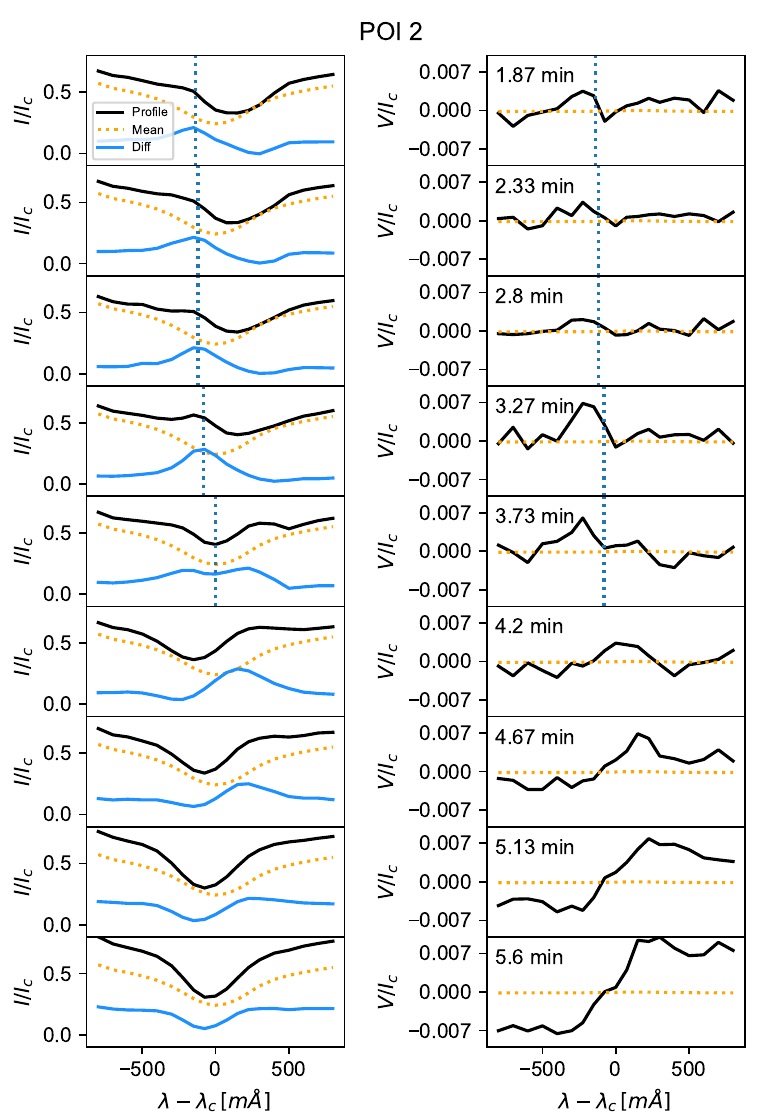}
\caption{Time sequences of the Stokes-$I$ and $V$ profiles of the \ion{Ca}{II} line at $8542.1\,\AA$ normalised to the continuum intensity for two points of interest (POI\,1 and POI\,2) in the close vicinity of the MBP: POI\,1 (left plots): black $\times$-sign in the first row of Fig.\,\ref{Fig5} and POI\,2 (right plots): black $+$-sign in first row of Fig.\,\ref{Fig5}. The actual line profiles are shown in black, the average profiles over the full FOV and over time are shown in orange, and the blue lines correspond to the difference between the black and orange profiles. Dotted blue vertical lines before $t=3.73\,\text{min}$ mark the position of the maximum value of the Stokes-$I$ emission excess for each profile (maximum of the blue curve). Dotted blue vertical lines after $t=3.73\,\text{min}$ mark the position of the Stokes-$V$ zero-crossing for the double-lobe signatures in the POI\,1 profiles. The sequences start at $t=1.87\,\text{min}$ and have a cadence of 28 seconds.}
\label{Fig6}
\end{figure*}

In Sect.\,\ref{SIR}, we distinguish between ascending and descending flows within the magnetic flux concentration in the photosphere. The difference between upflow and downflow increases with height in the atmosphere, reminiscent of a steepening wave that may develop into a shock wave when travelling upward into the chromosphere. In such a scenario, we expect an observable signature at chromospheric heights. To this end, in this subsection, we inspect the flow dynamics and emission signatures within and close to the magnetic flux concentration in the upper photosphere and the low and middle chromosphere based on the analysis of the \ion{Ca}{ii} line spectra.

Figure \ref{Fig5} summarises the plasma dynamics within the magnetic flux concentration and its surroundings at the photosphere and lower chromosphere based on several quantities derived from the profiles of the \ion{Ca}{II} line at $8542.1\,\AA$ and the \ion{Ca}{II} K line core. The time advances from left to the right from $t=0.93\,\text{min}$ to $t=5.6\,\text{min}$ with a cadence of $56\,$s, covering the last stage of the vortical drain phase and the complete upward-flow phase according to the definition in Sect.\,\ref{SIR}. The FOV of $8^{\prime\prime} \times 8^{\prime\prime}$ is centred on the magnetic flux concentration. 

The first and second rows of Fig.\,\ref{Fig5} show maps of the time evolution of the bisector velocity for the \ion{Ca}{II} line at $8542.1\,\AA$ at line depths of 70\% and 40\%, respectively. These percentages indicate the depth at which the bisector is taken with respect to the full line depth. At these line depths, we obtain information on the atmospheric conditions in the upper photosphere and the lower chromosphere from 500 km to 900 km. The maps in the top row include magnified views of a $1.5^{\prime\prime} \times 1.5^{\prime\prime}$ region centred on the MBP. From the initial minute of evolution, we recognise an annular region of downflows surrounding the MBP area. This region assumes a more asymmetric shape after one minute. These persistent downflows start at $t=0.93\,\text{min}$ and stop at $t=3.27\,\text{min}$, reaching maximum bisector velocities of $12\unit{km\,\text{s}^{-1}}$ at $t = 2.8\,\text{min}$. We associate this time range with the chromospheric drain phase, which is delayed by a minute compared to the vortical drain phase in the photosphere, discussed in Sect.\,\ref{SIR}. \cite{Fischer2009} observed in their statistics of convective collapse events also a downflow in the high photosphere during the evacuation process. Our results indicate that the downflow can take place even higher in the atmosphere, namely, in the chromosphere. 

Starting at $t=3.73\,\text{min}$, a sudden release of upward-flowing plasma is observed in the bisector velocity maps of Fig.\,\ref{Fig5}. The upflows occur in the surroundings of the MBP with a maximum bisector velocity of $-12\unit{km\,\text{s}^{-1}}$. The up-flowing plasma seems to grow in a clockwise sense during the next three minutes. The time frames, $t=3.73\,\text{min}$, $t=4.67\,\text{min}$ until $t=5.6\,\text{min}$ (see Fig.\,\ref{Fig5}), show an expansion of the upflow region initially characterised by a compact spiral shape, followed by the formation of a comma-shape pattern going from the central-right to the lower region of the FOV (best seen in the emission strength; see fifth row of Fig.\,\ref{Fig5}). We identified the same delay of one minute between the upward-flow phase in the photosphere and the upward-flow phase in the chromosphere. 

Based on a k-means classification of Stokes-$I$ profiles of the \ion{Ca}{II} line at $8542.1\,\AA$ for pixels within the $1.5^{\prime\prime} \times 1.5^{\prime\prime}$ region, we detect extended zones surrounding the MBP area that show conspicuous intensity excess as in the form of 'humps' in the blue and red wings of the line profile at different moments of the evolution of the event. In particular, profiles characterised by a hump in the blue wing and a redshifted core (red contours in the magnified views in the first column of Fig.\,\ref{Fig5}) are prevalent in the chromospheric drain phase, while profiles characterised by a hump in the red wing and a blueshifted core (blue contours in the magnified views in the first column of Fig.\,\ref{Fig5}) dominate after $t=3.73\,\text{min}$. We highlight two fixed points of interest (POIs; marked with a $\times$-sign and a $+$-sign in the magnified views in the first row of Fig.\,\ref{Fig5}) that exhibit the characteristic profiles of the selected groups and that show these conspicuous humps. The Stokes-$I$ and $V$ profiles in the POIs are shown in Fig.\,\ref{Fig6} and analysed in Sect.\,\ref{shock}. 

Representing the even higher layers of the atmosphere, the third row of Fig.\,\ref{Fig5} shows the temporal sequence of the normalised core intensity of the \ion{Ca}{ii} line at $8542.1\,\AA$ in logarithmic scale. Areas of enhanced intensity stay persistently bright in all frames, except for the last one, and a fibrilar spiral pattern is visible surrounding the central part of the FOV, connecting it with external regions. Areas with the highest intensities correspond to locations of spectral profiles with a decrease in line depth of $43\%$ of the average line profile, indicating line weakening. At the location of the MBP, we observe a persistent and extended brightening during the chromospheric drain phase that moves towards the upper edge of the MBP at $t=3.73\,\text{min}$ (see position $(x,y)=(4.0^{\prime\prime},4.5^{\prime\prime})$) and seems to dissipate later on. To search for spatial connectivity with external magnetic flux concentrations, we superposed the signed total circular polarisation (TCP) for values greater than 2.5$\sigma_{p}$ of the \ion{Fe}{i} line at $6173.3\,\AA$ on the maps of the third row of Fig.\,\ref{Fig5}. We distinguish small magnetic flux concentrations of equal polarity (orange contours) and inverse polarity (green contours) with respect to the central magnetic flux concentration associated with the MBP. They occur cospatial with particularly bright regions of the fibrilar eddy pattern. For instance, there are inverse-polarity patches prevalent during the entire sequence located around $(x_1,y_1)=(4.2^{\prime\prime},1.2^{\prime\prime})$ and $(x_2,y_2)=(1.3^{\prime\prime},5.7^{\prime\prime})$.

The fourth row of Fig.\,\ref{Fig5} shows the temporal evolution of the K-index (as defined in Table \ref{t1}) in logarithmic scale. Bright areas in these maps indicate emission in the low and middle chromosphere, testifying to regions with a temperature stratification different from the mean stratification, given the important contribution of the \ion{Ca}{ii} K line to chromospheric radiative losses \citep{Rezaei2007_2}. Superposed on the K-index maps are contours indicating regions where the $K_{2V}$ band (red contours) and $K_{2R}$ band (blue contours) have an intensity excess above 0.2$I_c$. We notice that the intensity excess in the $K_2$ spectral components associated with the MBP evolves according to the defined flow phases explained as follows.

During the drain phase, the intensity excess of the $K_{2V}$ band (red contours) increases from $t = 0.93\,\text{min}$ to $t = 2.8\,\text{min}$, when it reaches its maximum value and for the most part encloses the MBP. As demonstrated by \cite{Carlsson1997}, such enhanced emission in the $K_{2V}$ band, also called chromospheric bright grain, stems from an acoustic wave that steepens into a shock in the middle chromosphere. Chromospheric bright grains can be explained by the opacity window effect given by strong differential flows throughout the atmosphere \citep[\& references therein]{Bose2019, Mathur2022}. In this scenario, the shock propagates into a down-flowing upper atmosphere leading to a reduced contribution to the absorption on the blue side of the line above the shock. Together with the blue-shifted opacity due to the enhanced density of the post-shock material, this process drives the $K_{2V}$ band in emission \citep{Carlsson1997}. During the upward-flow phase, the $K_{2V}$ band intensity excess disappears giving way to a small area of $K_{2R}$ band intensity excess within the MBP area (blue contours), and that persists in the following minutes. This signature confirms the transition in the flow direction within the atmosphere, shifting from downflow to upflow.

The emission strength and $K_{2}$-asymmetry, as defined in Table\,\ref{t1}, are shown in the fifth and sixth row of Fig.\,\ref{Fig5}, respectively. The panels presented in the fifth row of Fig.\,\ref{Fig5} reveal the existence of a persistent chromospheric vortical structure centred on the MBP. This pattern strongly resembles the chromospheric swirls observed by \cite{Wedemeyer2009}. The temporal evolution of these quantities agrees with the scenario of a swirling structure anchored to the MBP, which is persistent in time. The high spatial resolution of these maps reveals the fine structure of the swirl, characterised by multiple fibrilar spiral arms, localised in regions of downward and upward plasma flows. We include a time-lapse animation of Fig.\,\ref{Fig5} available online to observe the complete evolution of the chromospheric swirl at the different chromospheric observables.

In summary, we find that up and downflows increase in amplitude with height, reaching velocities in the chromosphere of $\pm 12\unit{km\,\text{s}^{-1}}$ but delayed by 1 min relative to the respective photospheric flow phases. The \ion{Ca}{II} infrared line and the $K_{2V}$ and $K_{2R}$ passbands show humps and intensity excesses typical of chromospheric bright grains and maps of the line core of the \ion{Ca}{II} infrared line and the \ion{Ca}{II} K index show a chromospheric bright grain and a fibril spiral pattern typical of chromospheric swirls.

\subsection{Evidence of an acoustic shock wave}
\label{shock}

\begin{figure*}
\centering
\includegraphics[width=\textwidth]{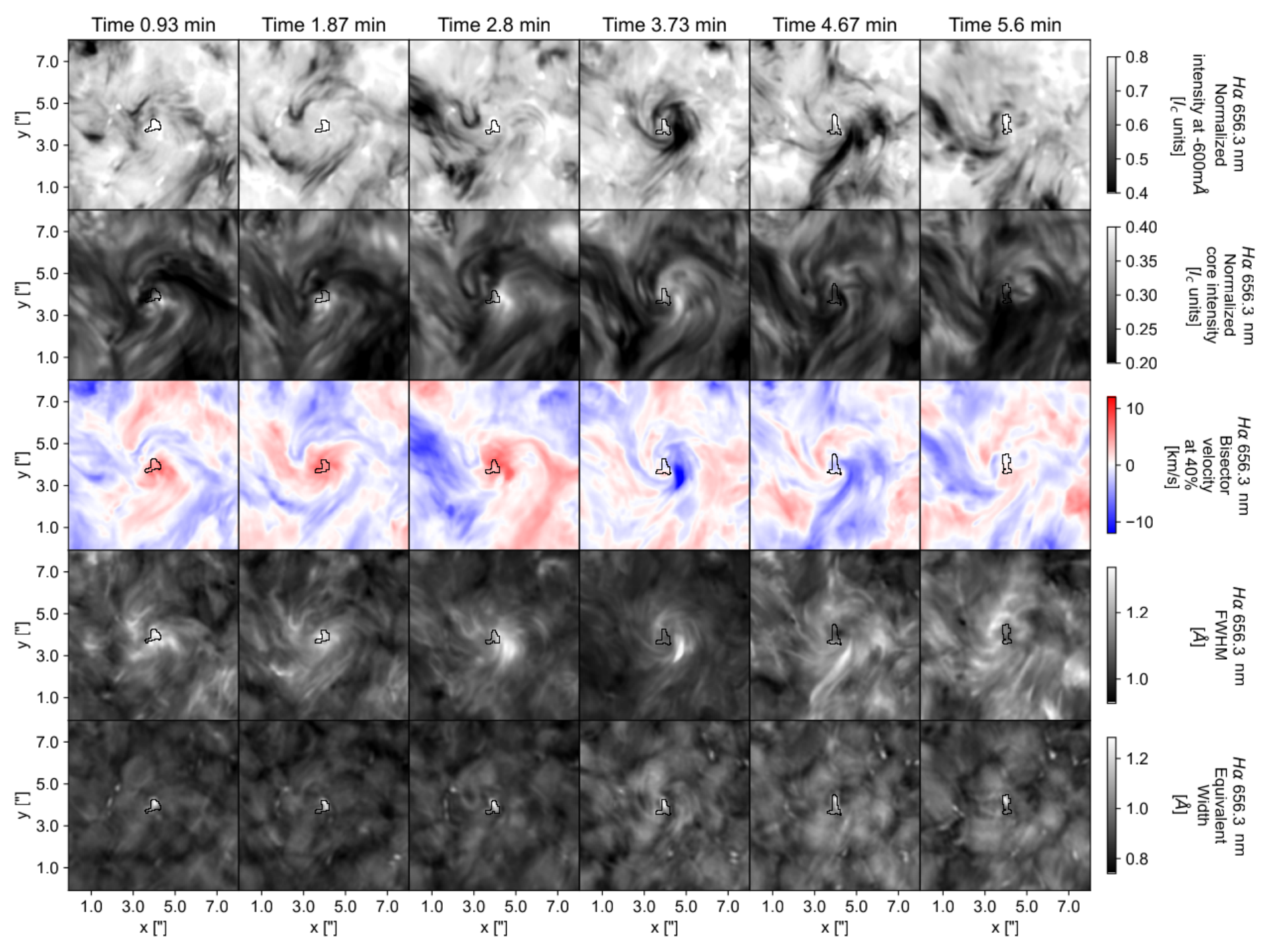}
\caption{Time series of maps of various properties of the spectral line H$\alpha$ at one minute cadence. First row: normalised intensity in the blue wing of H$\alpha$ at -600\,m$\AA$ maps from the line core. Second row: normalised intensity maps in the line core. Third row: Bisector velocity at 70\% depth of the H$\alpha$ line core Fourth row: Bisector velocity at 40\% depth of the H$\alpha$ line core. Fifth and Sixth row: FWHM and EW of H$\alpha$, respectively. All maps include a central black contour to show the location, scale, and evolution of the MBP. An animation is available online.}
\label{Fig8}
\end{figure*}

The extended zones of conspicuous intensity excess found in the blue wing of the \ion{Ca}{ii} line at $8542.1\,\AA$ (red contours in the insets of the Fig.\,\ref{Fig5}, first row) might indicate the existence of a shock wave travelling from the photosphere to the chromosphere. The spectral signatures found in this area expose some characteristics of the propagation of such a perturbation. In this subsection, we study the time series of Stokes profiles of the infrared Ca II line in search of propagating disturbances in the chromosphere caused by the MBP.

Figure \ref{Fig6} summarises the time evolution of the Stokes $I$ and $V$ mean profiles for a $2\times2$ pixel box centred on the two points of interest (POIs) indicated in the zoomed panels in the first row of Fig.\,\ref{Fig5}. The POI\,1 is located in the upper left corner close to the MBP (black $\times$-sign), where we detected the strongest downflows. The POI\,2 is located in the upper edge close to the MBP (black $+$-sign), in the region with the greatest excess in line core brightness. 

In both flow phases and both POIs, we observed intensity excesses as humps in the Stokes-$I$ profile: 1) in the blue wing during the drain phase and 2) in the red wing during the upward-flow phase. The hump in the blue-wing is first detected at $t=2.33\,\text{min}$ in POI\,1, reaching its maximum intensity excess at $t=2.8\,\text{min}$. The hump is centred at $0.1\,\AA$ from the central wavelength, corresponding to a Doppler velocity of $\Delta v = -3.5\unit{km\,\text{s}^{-1}}$. In the case of POI\,2, a less intense but visible blue-wing hump appears at $t=3.27\,\text{min}$. We detect a time lag of about 28 seconds from the blue-wing hump at $t=2.8\,\text{min}$ in POI\,1 to its appearance in POI\,2, which suggests a transversal clockwise propagation of the disturbance with a velocity of about $14\unit{km\,\text{s}^{-1}}$. At $t=3.73\,\text{min}$, the profile in POI\,2 exhibits an enhanced core intensity compared to the average profile, and a faint red-wing hump appears, potentially indicating the arrival of a shock wave at the formation height of the line core (i.e. weakening of the absorption line and no apparent Doppler shift). At $t=4.2\,\text{min}$ and $t=4.67\,\text{min}$ both POIs exhibit overall blueshifted line profiles, accompanied by the appearance of the hump in the red-wing. For profiles observed at POI\,1, the red-wing hump is located around $+0.3\,\AA$, corresponding to a Doppler velocity of $\Delta v = 10\unit{km\,\text{s}^{-1}}$. 

These intensity excesses in the blue and red wings of the Stokes-$I$ profiles are accompanied by characteristic signals in the corresponding Stokes-$V$ profiles, again with different spectral properties depending on the flow phase. In some cases, these signals reach amplitudes of around $3\sigma_{p}$. For instance, during the drain phase at $t=2.8\,\text{min}$ in POI\,1 and at $t=3.27\,\text{min}$ in POI\,2, a single-lobe Stokes-$V$ profile appears at the wavelength position of the blue-wing hump of Stokes-$I$. During the upward-flow phase at $t=4.2\,\text{min}$ and $t=4.67\,\text{min}$ in POI\,1, a typical double-lobe Stokes-$V$ profile appears with the wavelength of the zero-crossing coinciding with the red-wing hump of Stokes-$I$. Nonetheless, we notice that the double-lobe signatures associated with the red-wing hump correspond to an anomalous signal in Stokes $V$ since it has positive polarity while the magnetic flux concentration studied has negative polarity. These anomalous signals have a magnetic flux density of around $100\unit{G}$ as inferred by the weak field approximation \citep{Centeno2018} of the \ion{Ca}{ii} line. For comparison, pixels inside the MBP exhibit values of $\approx 250\unit{G}$, also inferred by the weak field approximation of this line. Considering that the Stokes-$V$ profile with the inverted polarity is in the wavelength position of the red-wing hump in the Stokes-$I$ profile, this profile can be interpreted as being in emission, which flips the polarity.

The in-emission Stokes-$V$ component in the red wing of the line is present in pixels near the boundary of the MBP as the only signal in Stokes $V$ above the noise level. On the boundary, it occurs in combination with a normal double-lobe profile with negative polarity centred on the Stokes-$I$ line core. Similar signatures of such anomalous circular polarisation profiles have been already studied in sunspot umbrae as a manifestation of magneto-acoustic shocks, commonly known as umbral flashes \citep[and references therein]{Socas_Navarro2000, delaCruzRodriguez2013}. Anomalous Stokes-$V$ profiles were also shown to occur as a consequence of a shock front in numerical simulations of magnetic flux concentrations \citep{Steiner1998}. Based on the k-means classification, pixels with anomalous Stokes-$V$ profiles having the inverse polarity form a semi-closed ring around the magnetic flux concentration during the upward flow phase (blue contours in the first row of Fig.\,\ref{Fig5}). This suggests the existence of an extended zone of profiles with a Stokes-$V$ component which is in-emission in the red wing of the line.

Summing up, we found humps in the blue and red wing of Stokes-$I$ during the drain phase and the upward-flow phase, respectively, which can be interpreted to be due to a vertically propagating shock front that seems to travel also in the transverse direction relative to the line of sight following a spiral pattern in the chromospheric layers of the MBP. The corresponding Stokes-$V$ profiles are partially in emission having inverse polarity or appear as one-lobed profiles as a consequence of the strong dynamics.

Appendix \ref{wavelets} further deepens this analysis by performing a wavelet power analysis, which reveals excess power near the cut-off frequency at the location of the chromospheric bright grain. This supports the shock-wave hypothesis, as such an oscillatory signature is expected in association with a shock wave.

\subsection{Chromospheric jet}
\label{RBE}

\begin{figure}
\centering
\includegraphics[width=9cm]{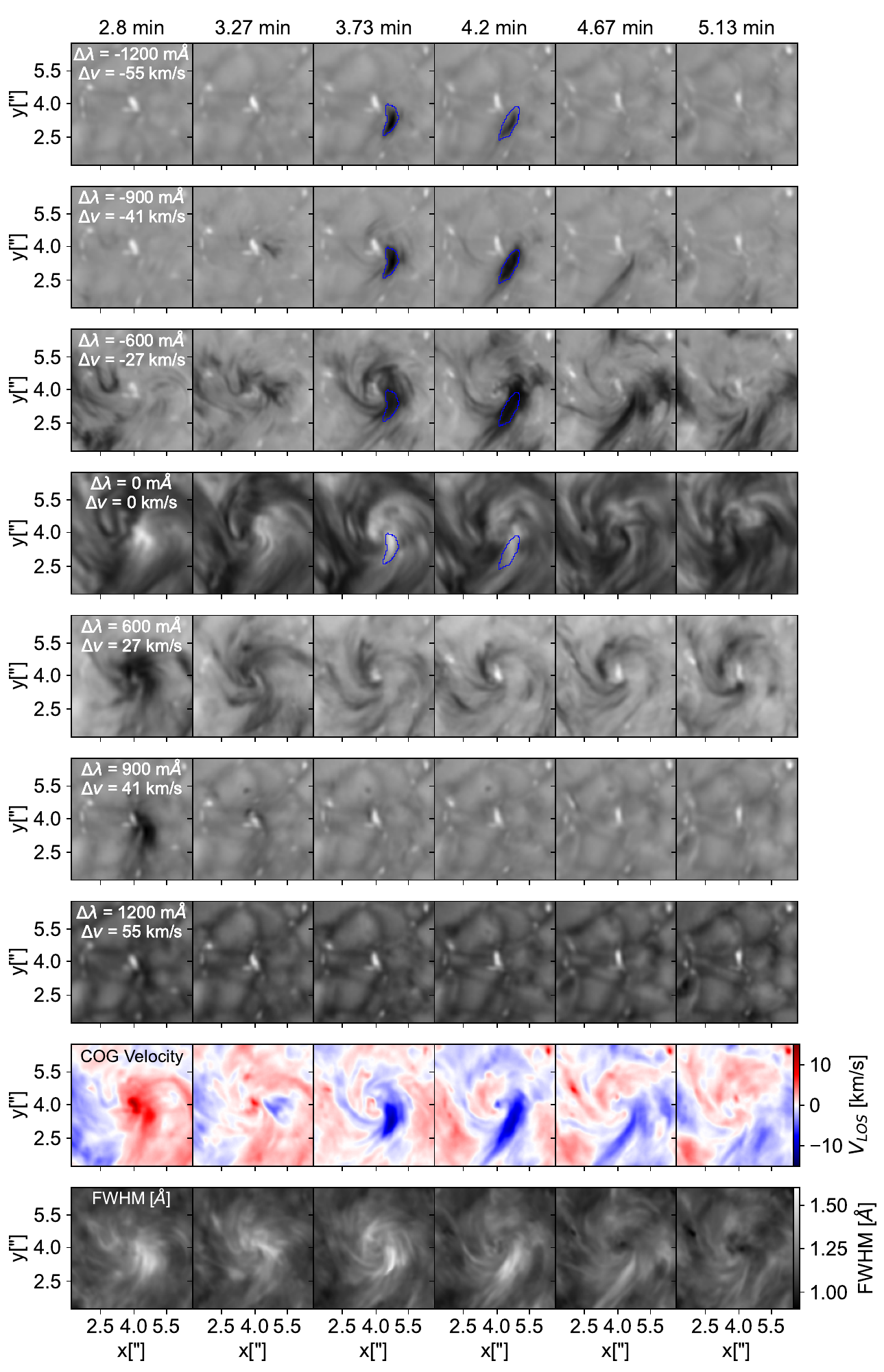}
\caption{Detailed cutouts showing the evolution of H$\alpha$ from $t=2.8\,\text{min}$ to $t=5.13\,\text{min}$ with 28.2 seconds cadence where the chromospheric jet appears in different wavelength positions, which are indicated within the panels of the first column. The second but last and last row correspond to the LOS velocity based on the Centre of Gravity (COG) method and the FWHM, respectively. The blue contours mark the region of the dark feature at the line position $\Delta \lambda$ = -900 m$\AA$.}
\label{Fig9}
\end{figure}

\begin{figure}
\centering
\includegraphics[width=9cm]{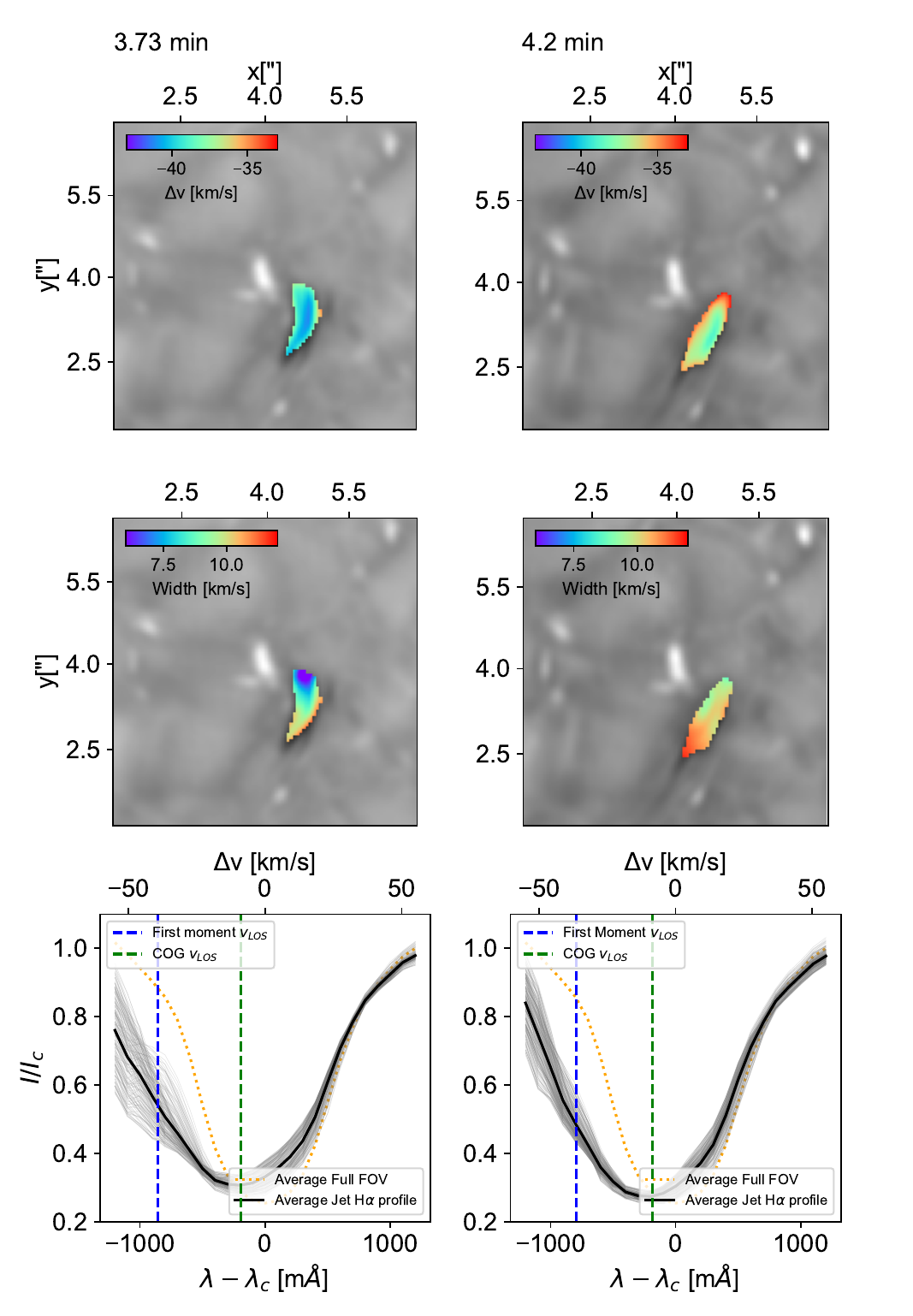}
\caption{Spectral properties of the H$\alpha$ profiles within the chromospheric jet for two time steps when it is visible. The background image corresponds to the intensity at the position $\Delta \lambda$ = -1200 m$\AA$ from the H$\alpha$ core. First and second row: Doppler velocity $\Delta v$ and line width, respectively. Third row: Full H$\alpha$ profiles for pixels within the dark feature (grey profiles), it is the correspondent mean profile (black profile) and the H$\alpha$ average profile for the full FOV as reference (orange dotted profile). The blue dashed line marks the Doppler velocity (first moment) of the separated blueshifted component calculated for the average profile of the jet, and the green dashed line marks the COG velocity calculated for the average profile of the jet.}
\label{Fig10}
\end{figure}

\begin{figure*}
\centering
\includegraphics[width=\textwidth]{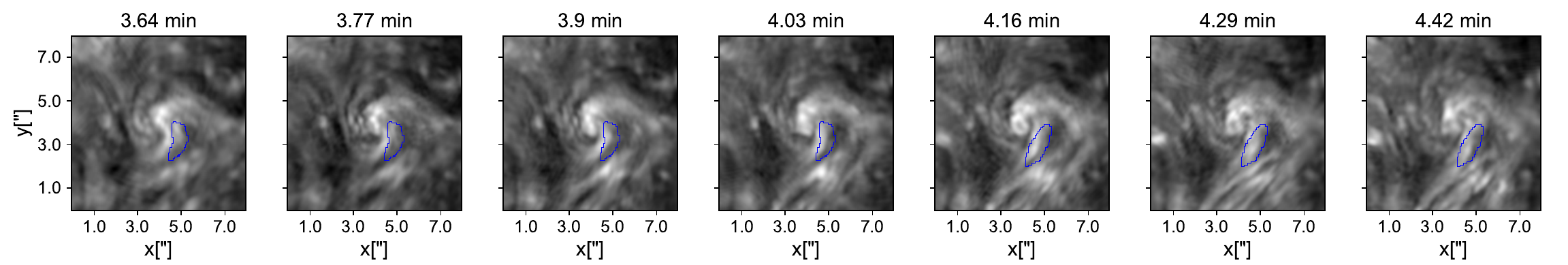}
\includegraphics[width=\textwidth]{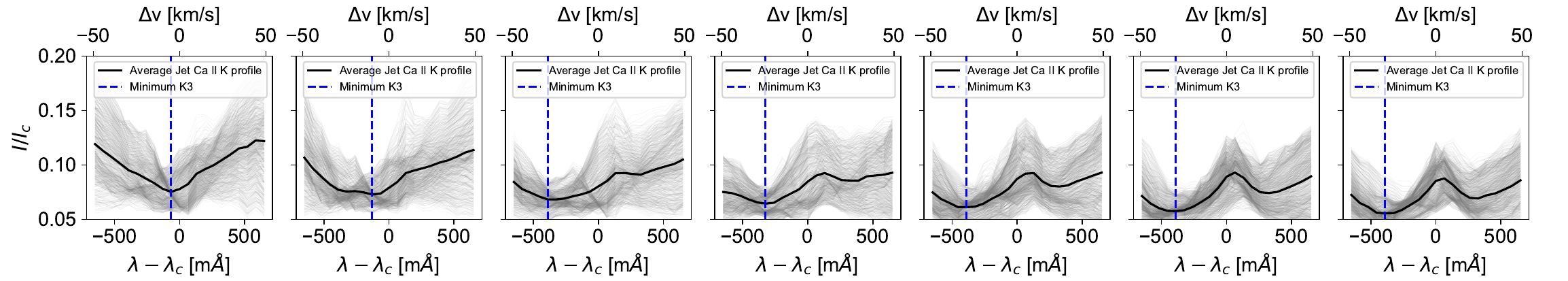}
\caption{Full cadence observation of the \ion{Ca}{II} K spectra in the line core during the detection of the chromospheric jet between $t=3.64\,\text{min}$ to $t=4.42\,\text{min}$. Top row: Maps of the intensity in the fixed wavelength at $3933.7\,\AA$. Blue contours mark the location of the jet according to the H$\alpha$ observation. Bottom row: The corresponding spectral profiles in each time step of pixels within the jet (grey profiles) and their mean profile (black profile). The blue, vertical, dashed line marks the minimum value of the line core.}
\label{Fig11}
\end{figure*}

In this subsection, we examine the time sequence of maps of various quantities derived from the H$\alpha$ spectral line, hence, again the chromospheric consequences of the initial MBP and vortex. Figure \ref{Fig8} shows several quantities computed from the H$\alpha$ spectra from $t=0.93\,\text{min}$ until $t=5.6\,\text{min}$. It confirms the persistent spiral morphology centred on the MBP as can be seen from the intensity images taken at $-600 \,\text{m\AA}$ from the H$\alpha$ line core (first row of Fig.\,\ref{Fig8}) and from the intensity images in the line core (second row of Fig.\,\ref{Fig8}), both normalised to the continuum. The evolving swirling structure extends across the full FOV having an average diameter of $6^{\prime\prime}$, while the central spiral-shaped region, clearly resolved at $t=3.73\,\text{min}$, has a diameter of $\approx 3^{\prime\prime}$. From the bisector velocity at 40$\%$ (third row of Fig.\,\ref{Fig8}), we find that the downflows reach middle chromospheric heights during the drain phase, having a maximum downflow velocity of $9\unit{km\,\text{s}^{-1}}$ at $t=2.8\,\text{min}$. These strong downflows correlate with increased values of FWHM (fourth row of Fig.\,\ref{Fig8}) but without appreciable changes in the EW (fifth row of Fig.\,\ref{Fig8}). The temporal evolution of the chromospheric swirl observed in H$\alpha$ is included as a time-lapse animation of Fig.\,\ref{Fig8} available online.

At $t=3.73\,\text{min}$ a dark feature in the blue wing of the H$\alpha$ line is observed within the chromospheric swirl. To explore the properties of this dark feature, we include Fig.\,\ref{Fig9}, showing the full cadence intensity maps at different wavelength positions of the H$\alpha$ line essentially during the upward-flow phase together with the LOS velocity and FWHM spatial distributions as reference. The LOS velocity is calculated using COG based on the approach of \cite{Uitenbroek2003} and the FWHM is calculated using the method described in Sect.\,\ref{methods}. The sequence spans from $t=2.8\,\text{min}$ to $t=5.13\,\text{min}$ for seven wavelength positions around the line centre. In the bluest wavelength position sampled, corresponding to a Doppler shift of $\Delta v = -55\unit{km\,\text{s}^{-1}}$ (first row), the dark feature is only observed at $t=3.73\,\text{min}$ and at $t=4.2\,\text{min}$ while in the other wavelength positions the swirl structure appears gradually when approaching the line core (fourth row). In wavelength positions close to the line core, the dark feature follows the spiral pattern of the chromospheric swirl and progressively stretches and fills the region below the main spiral arm at $t=4.67\,\text{min}$. 

The eighth and ninth rows of Fig.\,\ref{Fig9} show the LOS velocity and FWHM distributions, respectively. The LOS velocity shows the evolution of the swirl between upward and downward flows, going from fast downflows with maximum velocities of $15\unit{km\,\text{s}^{-1}}$ at $t=2.8\,\text{min}$ to fast upflows with maximum velocities of $-15\unit{km\,\text{s}^{-1}}$ when the dark feature is observed. Based on its visual characteristics, we are most likely observing a fast and coherent bulk of plasma moving upwards along the swirl, i.e., a chromospheric jet. 

The chromospheric jet appears bright in the FWHM images having an average width of 1.6$\,\AA$. However, no visible changes occur in the EW maps depicted in the last row of Fig.\,\ref{Fig8}. To quantify marginal changes in the EW maps, we compute the spatial average of the EW across the entire FOV for each frame. Subsequently, we compare each spatial average with its temporal average value derived from the complete time series. Our analysis reveals that the spatial average of the EW shows a 0.5\% increase at $t=3.73\,\text{min}$, relative to preceding frames. Such a rise in the EW during the upward flow phase evidences heating signatures.

We find that many of the visual and spectral properties of the present chromospheric jet are similar to those of a rapid blue-shift excursion \citep[RBE -][]{Langangen2008, RouppevanderVoort2009, Sekse2012, Bose2019}. We determine that the present jet has a lifetime of around 60 seconds, is aligned with the swirl structure, i.e., it follows a logarithmic spiral path, has a maximum length of around $1.1\,\unit{Mm}$, and moves with an apparent transversal velocity of $11\unit{km\,\text{s}^{-1}}$, in agreement with the values reported by \cite{Sekse2012} for RBEs.

Although the jet is localised in a region of an upward flow, its characteristics differ from the other upflow regions within the swirl as it develops highly asymmetric profiles toward the blue. Profiles from the jet have a blue-shifted core which is not typically observed in RBEs, however, strong asymmetry observed in the blue wing might be related to a separated blue-shifted component. To estimate the properties of the blue-shifted component, we compute its Doppler velocity and line width using the method proposed by \cite{RouppevanderVoort2009} for profiles within the jet. The author states that such an estimation is rather crude, thus the values provide only rough estimations for the actual quantities. 

The first and second rows of Fig.\,\ref{Fig10} show the distribution of the Doppler velocity ($\Delta v$) and line width (Width), respectively, within the jet for the time steps in which it becomes visible in the first row of Fig.\,\ref{Fig9} as computed according the method of \cite{RouppevanderVoort2009}. At $t=3.73\,\text{min}$ (left column), the jet appears as a compact structure with velocities around $-40\unit{km\,\text{s}^{-1}}$ along the length of the streak, which follows the swirl pattern. Its line width varies systematically along the length, having small widths at the footpoint of the jet and near the MBP widening up on the other end. Later, at $t=4.2\,\text{min}$ (right column), the jet stretches and slows down, reaching velocities of around $-35\unit{km\,\text{s}^{-1}}$, but remaining compact. The systematic variation of the line width along the length also remains but with an overall increase with respect to the previous time step. 

We identified variations in the asymmetry of the profiles in the jet during the time steps considered. The third row of Fig.\,\ref{Fig10} shows the distribution of H$\alpha$ profiles of the pixels within the jet (grey curves) and their mean (black curves) for $t=3.73\,\text{min}$ (left column) and $t=4.2\,\text{min}$ (right column). As a reference, the spatial and temporal average profile of the full FOV is included in each panel (dotted orange curves). In these two time steps the H$\alpha$ profiles behave differently. On the blue side, the profiles are highly dispersed at $t=3.73\,\text{min}$, indicating a fine structure inside the jet. In contrast, the profile distribution looks more uniform at $t=4.2\,\text{min}$ characterising an atmosphere with similar thermodynamic properties. The COG velocities calculated for the mean H$\alpha$ profile do not change dramatically during the time steps considered. They remain within the range of $-8$ to $-9\unit{km\,\text{s}^{-1}}$ (green vertical dashed lines in the third row of Fig.\,\ref{Fig10}). 

The LOS velocities calculated with the COG method and the LOS velocities calculated with the bisector method around 40\% of the line depth share similar values, ranging between $-5$ to $-15\unit{km\,\text{s}^{-1}}$. This can be expected as both these methods account for the overall displacement of the entire line, yielding a mean velocity of the atmosphere that is sampled by the H$\alpha$ line. In contrast, the LOS velocities computed with the Doppler shift method proposed by \cite{RouppevanderVoort2009} are larger than these mean values, because this method is sensitive to asymmetries of the blue wing of the line, thus providing velocities of the secondary and blueshifted component, which is blended with the main component of the line. Hence, the chromospheric jet moves much faster than the upward flows in the chromospheric swirl.

Typical RBEs can be also observed in \ion{Ca}{II} spectra \citep{Sekse2012,Bose2019}, hence, we inspected the \ion{Ca}{ii} at $8542.1\,\AA$ and \ion{Ca}{ii} K spectral lines for possible signatures of the chromospheric jet at their formation heights. The upward flow phase is certainly present in both lines, as shown in Fig.\,\ref{Fig5} and described in Sect.\,\ref{swirl}; therefore, the line profiles with a blue-shifted core are recognised within and near the jet. Upon inspecting in greater detail, we notice that only a few pixels near the footprint of the jet show appreciable asymmetries in the blue wing of the \ion{Ca}{ii} line at $8542.1\,\AA$, and no compact dark feature is present in the vicinity of the chromospheric jet. Thus, there is no clear counterpart of the jet in the upper photosphere.

Unlike the \ion{Ca}{ii} line at $8542.1\,\AA$, the \ion{Ca}{ii} K line core samples heights in the chromosphere where one can most likely identify counterparts of the jet observed in H$\alpha$. Taking advantage of the higher cadence and higher spatial resolution of the \ion{Ca}{ii} K records, we studied the detailed evolution of the \ion{Ca}{ii} K spectra during the lifetime of the jet. The first row of Fig.\,\ref{Fig11} shows the sequence of seven frames between $t=3.64\,\text{min}$ to $t=4.42\,\text{min}$ covering the full lifetime of the jet. The images display the intensity normalised to the continuum at a fixed wavelength position at $3933.7\,\AA$ and the corresponding location and area of the jet observed in H$\alpha$ (blue contours). The swirl structure is enhanced in all frames, but no particular feature characterises the jet. Analysing all spectral profiles within the jet, shown in the second row of Fig.\,\ref{Fig11}, we find a blueshift of the $K_3$ component (minimum value of the line core) of the line starting at $t=3.77\,\text{min}$, just a few seconds after the initial detection of the jet in H$\alpha$. During the next minute until $t=4.42\,\text{min}$, a progressive enhancement of the $K_{2R}$ component was observed. According to \cite{Bose2019}, these profiles exhibit the typical shape of the \ion{Ca}{II} K spectra in RBEs, in which the absorption signatures are dominated by opacity shifts caused by steepened velocity gradients. In the present case as well, the $K_3$ component is blueshifted, causing the $K_{2V}$ component to be suppressed and the $K_{2R}$ component to be enhanced. 

In summary, maps of H$\alpha$, for instance, at -600 m$\AA$ from the line core, confirm the persistent spiral pattern centred on the MBP already seen from maps of the \ion{Ca}{II} K line with downflows seen to reach the middle chromosphere in the drain phase. Examining maps taken at various wavelength positions in the blue wing of H$\alpha$ reveals a dark feature that can be identified as a chromospheric jet that shares several characteristics with a rapid blue excursion, RBE. Different from typical RBEs, the H$\alpha$ line core of the present jet is slightly shifted to the blue but like typical RBEs, signatures of the jet are also visible from the blueshifts of the line core of \ion{Ca}{II} K.

\section{Summary and discussion}
\label{summary}

\begin{table*}[h!]
      \caption[]{Magnetic field strength and the LOS velocity ranges in each phase and height.}
         \label{t3}
         \[\begin{array}{p{0.32\linewidth}cccccc}
                        & \multicolumn{2}{c}{\text{Vortex drain phase}} & \multicolumn{2}{c}{\text{Upward-flow phase}} & \multicolumn{2}{c}{\text{Relaxation phase}} \\
            \noalign{\smallskip}
            \hline
            \noalign{\smallskip}
                        & |B| {[\unit{kG}]}  & |\text{v}_{LOS}| {[}\unit{km\,\text{s}^{-1}}{]}  & |B| {[\unit{kG}]}  & |\text{v}_{LOS}| {[}\unit{km\,\text{s}^{-1}}{]} & |B| {[\unit{kG}]} & |\text{v}_{LOS}| {[}\unit{km\,\text{s}^{-1}}{]} \\
            \noalign{\smallskip}
            \hline
            \hline
            \noalign{\smallskip}
            Photosphere             & 0.7 - 1.2    &  \sim 1                       & 1.0 - 1.6    & 2 - 3                  & 0.5 - 1.0   & \sim 1                      \\
            Low \& middle chromosphere & \sim 0.2^*         & 5 - 10                  & \sim 0.25^*         & 10 - 12                & \sim 0.17^*        & 0 - 3                      \\
            Upper chromosphere      & -            &  12 - 15                      & -            & 5 - 40                     & -           & 0 - 3                     \\
            
            \noalign{\smallskip}
            \hline
         \end{array}\]
         \footnotesize{$^*$ Estimation using the weak-field approximation.}
         \tablefoot{The vortex drain phase and the relaxation phase are dominated by downflows in contrast to the upward-flow phase.}
\end{table*}

\begin{figure*}
\centering
\includegraphics[width=\textwidth]{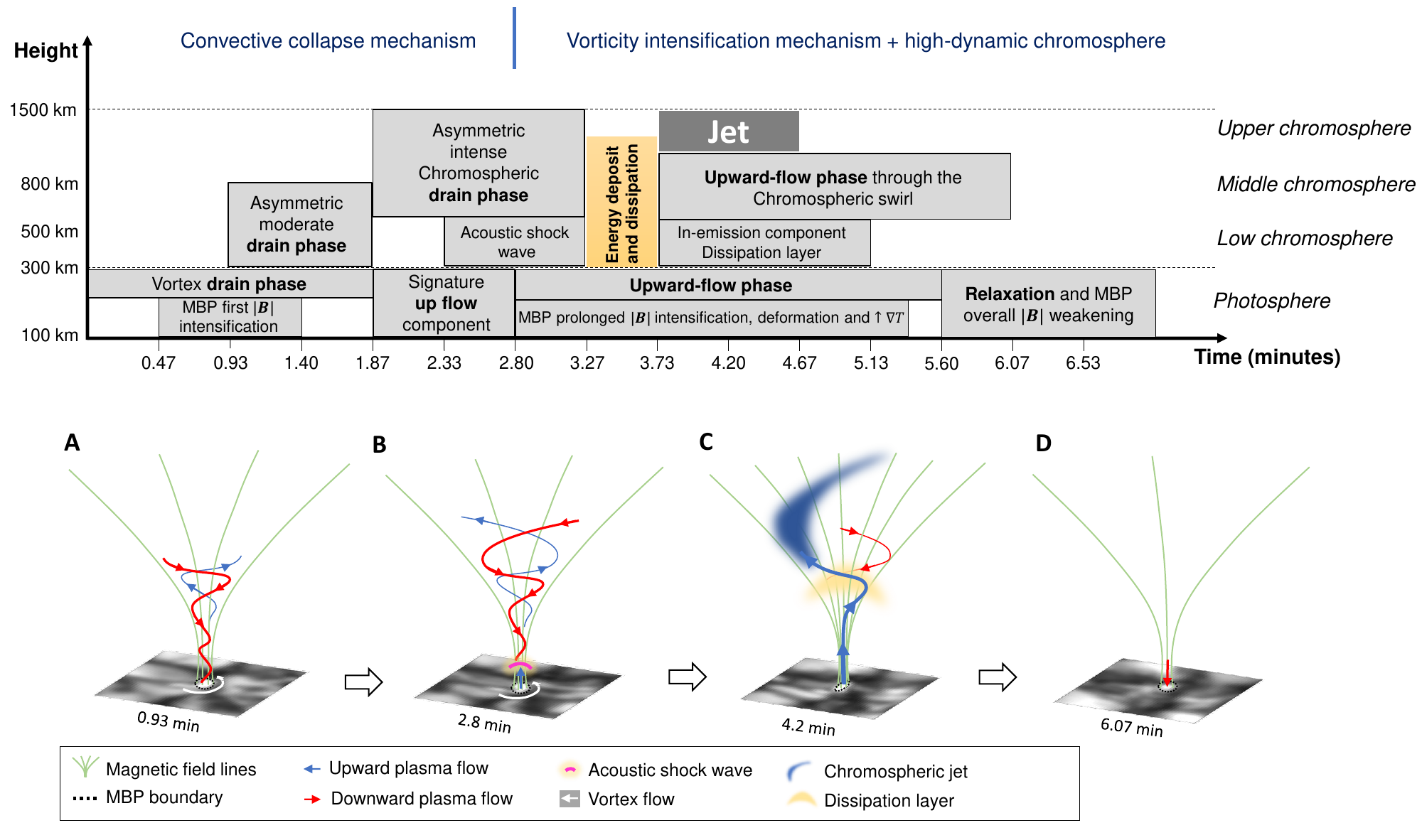}
\captionsetup{textfont=small}
\caption{Sketch and cartoons summarising the dynamics of the MBP interacting with a vortex flow attached to a chromospheric swirl and the release of an impulsive chromospheric jet. The detailed description is found in the Section \ref{summary}.}
\label{Fig12}
\end{figure*}

We present a complete and comprehensive description of the evolution of a magnetic flux concentration, classified as MBP, interacting with an intergranular vortex flow. In the initial stage of evolution, the vortex seems to alter the shape of magnetic flux concentration. Unlike the vortical flows studied by \cite{Bonet2008}, where several MBPs interact with one intergranular vortex, the present magnetic flux accumulation seems to have gathered all the neighbouring magnetic flux, forming a logarithmic spiral with one spiral arm. 

The magnetic flux concentration, with a maximum size of around $0.7^{\prime\prime}$ in the photosphere, harbours a fine-scale magnetic structure, in which the core of the MBP is affected by sequential intensification and weakening of the magnetic field. In particular, the observed prolonged magnetic field intensification behaves differently from the other two instances of intensification observed during the evolution of the MBP core. This may indicate different types of intensification. We identify the initial instance of magnetic field intensification of the MBP in agreement with the convective collapse scenario numerically simulated by \cite{Grossmann-Doerth1998} and observationally confirmed by \cite{BellotRubio2001}. Considering the vortical structure that we find within the MBP, we propose that the type of magnetic field amplification due to vorticity reported by \cite{Keys2020} from numerical simulations, might occur during the prolonged magnetic field intensification. Their simulation indicates that the vorticity in the MBP increases before the magnetic field amplification and then decreases when the magnetic field reaches its maximum value. This scenario may apply to the present observations, as the vortex appears at $t=1.4\,\text{min}$ and only exists previous to the intensification, namely, before $t=3.73\,\text{min}$.

The strong downflow within and in the surrounding of the MBP in the initial stage of the event entails a vortical flow, presumably by angular momentum conservation \citep[see e.g.][]{Nordlund1985}. Similar but at a bigger scale, \cite{Requerey2018} found that magnetic flux concentrations in the network are strongly affected by the evolution of a nearby vortex flow.

The magnetic flux concentration is associated with a persistent chromospheric swirl. It exhibits a spiral pattern with gas parcels moving upwards and downwards, guided by the host magnetic structure, which is anchored in the photospheric magnetic flux concentration (see panels A, B, and C in Fig.\ref{Fig12}). The present case exemplifies the spatial and temporal connectivity in a magnetised vortex, where there is an exchange of mass between the photosphere and chromosphere through an extended magnetic structure. Numerical simulations of magnetic and non-magnetic vortexes \citep[see e.g.][]{Shelyag2011, Moll2012} show the prevalence of vortices with vertically directed rotation axis in regions of magnetic flux accumulations. These cause the coupling of the near-surface to the upper layers, mediated by mainly vertical magnetic fields. This behaviour was also identified by \cite{Shetye2019} based on a statistical analysis of observations of chromospheric swirls.

To describe the spatial and temporal connectivity between the photosphere and the chromosphere, we summarise the dynamics of the event in terms of the temporal phases that occur time-shifted in different heights of the atmosphere. Table \ref{t3} summarises the magnetic field and LOS velocity ranges, evaluated with adequate methods for the different heights and the temporal phases. In addition, Fig.\ref{Fig12} shows a sketch summarising this evolution and a cartoon showing the physical scenario for four time-step characteristics of the phases of evolution (see panels A, B, and C in Fig.\ref{Fig12}), described below.

During the first two minutes of evolution, the drain phase takes place in the photosphere, whereby plasma motion is characterised by a downwards vortical flow within the magnetic flux concentration, and where the first magnetic field intensification occurs (see panel A in Fig.\ref{Fig12}). The downward flow leads to the evacuation of the magnetic flux concentration, characteristic of the intensification process due to the convective collapse \citep{Nagata2008}. In the chromosphere, the expanded magnetic structure seems to guide the plasma downwards along the chromospheric swirl that rotates in a counter-clockwise sense. 

From minute two onwards until $t=3.27\,\text{min}$, the delayed effect of the photospheric evacuation happens in the chromosphere (see panel B in Fig.\ref{Fig12}). Strong downflows surrounding the magnetic flux concentration are co-located with zones of enhanced brightening in the low chromosphere (see $t=2.8\,\text{min}$ in Fig.\,\ref{Fig5}). \cite{Moll2012} found in simulations of a magnetised quiet Sun region also a small-scale pattern of hot filaments tightly connected to vertically oriented vortices harbouring upflows and downflows. These high-temperature filaments were often associated with the shearing component of vorticity. With respect to the observations, the core of the magnetic flux concentration has a magnetic field strength below $1\,$kG, while the velocity gradient changes from a downflow to an upflow component, giving way to the propagation of an upward travelling perturbation and the start of the upward-flow phase in the photosphere (see panel B in Fig.\,\ref{Fig12}).

The manifestation of an upwardly travelling perturbation after the first magnetic field intensification agrees with the scenario proposed by \cite{BellotRubio1997}, where the plasma evacuation during the convective collapse leads to a rebound shock that propagates upwards. In the chromosphere, an emission component in the spectral line of \ion{Ca}{ii} at $8542.1\,\AA$ and the chromospheric bright grain announce the arrival of the perturbation as a quasi-acoustic slow wave guided by the magnetic field structure. It propagates upwards through an upper atmosphere that is downwardly flowing, similarly to the process described in \cite{Carlsson1997}. Moreover, this quasi-acoustic wave seems also to propagate transversely around the MBP in a clockwise sense while moving upwards, presumably reaching supersonic velocities. At this point, the MBP starts to exhibit the prolonged magnetic field intensification. 

From minute three onwards, several processes take place simultaneously during the upward-flow phase, with the major event being the release of a chromospheric jet accompanied by signs of shock dissipation (see panel C in Fig.\,\ref{Fig12}). In the photosphere, the MBP core harbours a magnetic field with a strength of above 1 $\unit{kG}$, reaching a maximum value of $1.6\,\unit{kG}$, and the upflows dominate the dynamic of the plasma, reaching the fastest upward speeds of $\approx 3 \unit{km\,\text{s}^{-1}}$. After $t=3.27\,\text{min}$, the photospheric vortex flow seems to disappear and a strong deformation of the MBP undergoes strong deformation through squeezing or bending. During this stage of evolution, it is not possible to be certain as to the cause of such deformation.  We note that it may be associated with the interaction between the last traces of the vortex flow with the magnetic flux concentration or caused by the gas pressure exerted by rising hot material in the close vicinity of the magnetic flux concentration. Such a process was shown in simulations of magnetic flux concentrations embedded in non-stationary convection motion in the surface layers of the convection zone \citep{Steiner1998}. Both dynamics can excite transversal waves propagating along magnetic structures and carrying mechanical energy into the upper layers of the solar atmosphere. 

In the chromosphere, the plasma flow changes directional sense from a downwards to upwards flow, following the structure of the swirl in a clockwise sense and exposing a chromospheric jet (see panel C in Fig.\,\ref{Fig12}). The morphology of the chromospheric swirl in the middle and upper chromosphere is best seen in \ion{Ca}{II} K and H$\alpha$. In agreement with the findings of \cite{Park2016}, we observe heating signatures in the form of relatively large values in the FWHM and EW of the H$\alpha$ profiles during the upward-flow phase, in particular after the jet release. 

The spectral characteristics of the jet share several similarities with typical RBEs observed near network regions \citep{RouppevanderVoort2009, Sekse2012, Bose2019}. In our case, the LOS velocity of the jet ranges from $-40\unit{km\,\text{s}^{-1}}$ to $-30\unit{km\,\text{s}^{-1}}$ inferred from the measurements of the Doppler shift of the separated blue-shifted component of the H$\alpha$ line and the Doppler shift of the minimum of the \ion{Ca}{ii} K line core during the jet detection. Simultaneously with the jet and after the arrival of the quasi-acoustic wave, the emission component in the red wing of the \ion{Ca}{ii} line at $8542.1\,\AA$ can also be seen in the polarimetric measurements. This is probably caused by a dissipation process associated with the jet release.

The oscillatory power excess at frequencies close to the cutoff frequency indicates locations of wave dissipation and chromospheric heating (explained in Appendix\,\ref{wavelets}). We infer that most of the power produced by the wave dynamics in the magnetic flux concentration is transferred from the inner part of the magnetic flux concentration to the surroundings following the spiral arms. Similar significant oscillating power below the cutoff frequency was previously found by \cite{Tziotziou2019} for a persistent chromospheric vortex flow, indicating the presence of fast kink waves. In our case, significant power is also present in frequencies above the cutoff frequency, probably produced by resonant interference or by the excitation of magneto-acoustic modes, such as torsional Alfvénic modes found in simulations \citep{Murawski2015, Battaglia2021}. In order to infer the effective torsion of the magnetic field of such magneto-acoustic modes, trustworthy measurements of the linear polarisation on such small scales are needed to effectively reconstruct the magnetic field vector.

A possible scenario that is comparable to the reported observations with the release of a high-speed jet is described by \cite{Kitiashvili2013} based on simulations. Their findings in terms of magnetic field strength ($1.2\,$kG in the photospheric layer), LOS velocity (upward velocity perturbation from  $6\unit{km\,\text{s}^{-1}}$ in the near-surface layers to more than $12\unit{km\,\text{s}^{-1}}$ above 700 km), and the triggering of shock waves agrees with our observations. In their case, the eruptions are driven in the photosphere by strong pressure gradients in the internal vortex structure, which later is further accelerated in the lower and middle chromosphere by Lorenz forces. They also identified the development of velocity perturbations associated with the vortex dynamics into shock waves. However, they did not report the propagation of a jet to the upper chromosphere, which, in our case, must be further accelerated with an impulse lasting less than a minute, taking into account the estimation of the maximum velocity of the jet. We suggest that the observed chromospheric jet is likely linked to strong pressure gradients in the photosphere that trigger the formation of acoustic shock waves.

From minute five onwards, the atmosphere slowly returns to its original downward flow direction and enters the relaxation phase, where the magnetic field strength decreases to below $1\,$kG in its overall structure (see panel D in Fig.\,\ref{Fig12}). The second magnetic field peak above $1\,$kG happens during this phase. This is possibly related to another convective collapse within the MBP core, which does not succeed in maintaining the stability of the magnetic flux concentration. Finally, the event concludes with its dissolution.

Our comprehensive analysis reveals the spatial and temporal connectivity between the photosphere and chromosphere in a vortical structure. This structure comprises a chromospheric swirl aligned with a single, small-scale magnetic field concentration. The observed characteristics strongly suggest that this event is a highly dynamic atmospheric vortex flow (AVF) or magnetic tornado, as described by \cite{Wedemeyer2012, Wedemeyer2014}. The event displays evidence of acoustic waves and triggered an impulsive chromospheric jet resembling an RBE.

\begin{acknowledgements}
The Swedish 1-m Solar Telescope is operated on the island of La Palma by the Institute for Solar Physics of Stockholm University in the Spanish Observatorio del Roque de los Muchachos of the Instituto de Astrofísica de Canarias. The Institute for Solar Physics is supported by a grant for research infrastructures of national importance from the Swedish Research Council (registration number 2021-00169). This project has received funding from the European Union’s Horizon 2020 research and innovation programme under grant agreement No 824135, the Trans-National Access Programme of SOLARNET. We thank Philip Lindner and Anjali J. Kaithakkal for recording and providing us with this dataset from their observational campaign. We also thank Oleksii Andriienko for performing the data reconstruction with SSTRED. We also thank Luc Rouppe van der Voort for his fruitful discussion and clarification regarding RBEs. Finally, we thank the referee for his/her comments and suggestions for improving our manuscript. S.M.D.C. and C.E.F. were funded by the Leibniz Association grant for the SAW-2018-KIS-2-QUEST project. The National Solar Observatory (NSO) is operated by the Association of Universities for Research in Astronomy, Inc. (AURA), under a cooperative agreement with the National Science Foundation.

\end{acknowledgements}

%
%

\bibliographystyle{aa}
\bibliography{references.bib}

\begin{appendix}

\section{Oscillatory signatures of the acoustic shock wave}
\label{wavelets}

In this Appendix, we further deepen the analysis of Sect.\,\ref{shock} by carrying out a wavelet power analysis. We search for oscillatory signals in connection with the propagation of the shock wave. The spectral signatures of the shock wave in the \ion{Ca}{ii} line at $8542.1\,\AA$ discussed in Sect.\,\ref{shock}, the detection of an enhanced emission in the $K_{2V}$ band of the \ion{Ca}{ii} K line core and the chromospheric bright grain in the location of the MBP at $t=2.8\,\text{min}$ suggest the presence of acoustic waves with frequencies close to the acoustic cutoff according to \cite{Carlsson1997}. To identify this possible oscillatory signature, we perform a wavelet analysis of the high cadence recordings of the \ion{Ca}{II} K line core. We investigate the intensity variation in the $K_{2V}$ band and in the minimum intensity of the line core, from now on call line-core minimum. We compute the two-dimensional (frequency/time) power spectra of the continuous wavelet transformation for all pixels of the calculated intensity maps ($K_{2V}$ band and line-core minimum) to inspect the characteristic acoustic modes in time using the standard Ricker wavelet \citep{Yanghua2014}. We consider two frequency ranges above the acoustic cutoff: 1) a medium-frequency range from $5.5\,$mHz to $8.5\,$mHz, and 2) a high-frequency range from $8.5\,$mHz to $16\,$mHz. The acoustic power of the intensity variations greater than $16\,$mHz is not taken into account as its signal is strongly affected by noise. The Nyquist frequency of the time series is $64\,$mHz. 

\begin{figure}
\includegraphics[width=9cm]{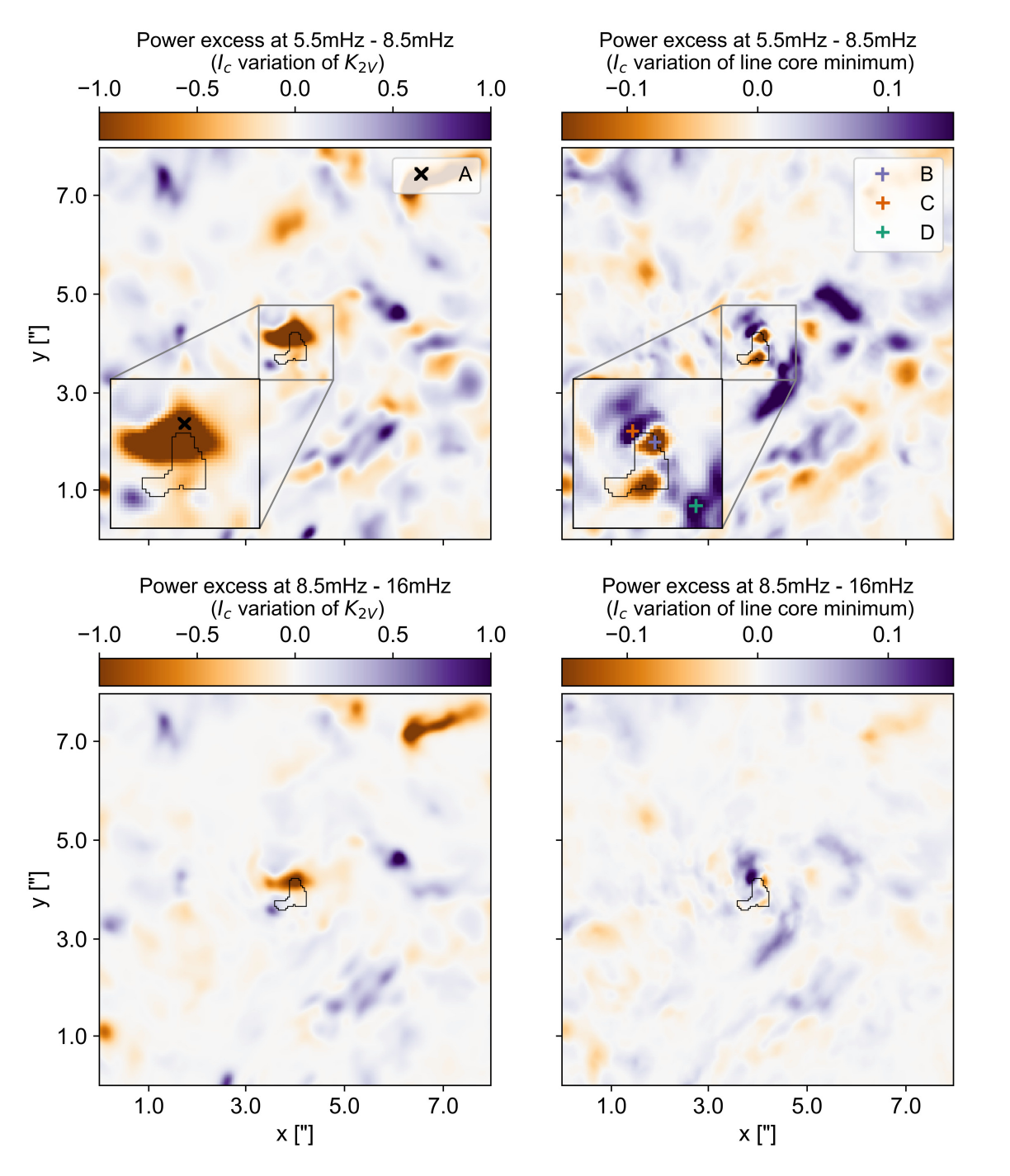}
\caption{Wave power excess in the medium-frequency range ($5.5\,$mHz to $8.5\,$mHz, top row) and in the high-frequency range ($8.5\,$mHz to $16\,$mHz, bottom row) of the intensity variation in the $K_{2V}$ band (left column) and line-core minimum (right column). Positive excess corresponds to larger values of power one minute after $t=3.27\,\text{min}$ and negative excess corresponds to larger values of power one minute before $t=3.27\,\text{min}$. The central black contour shows the location and scales of the MBP at $t=3.73\,\text{min}$.}
\label{Fig7}
\end{figure}

Based on the insight gained in Sect.\,\ref{shock}, we estimate that the shock wave detected in the \ion{Ca}{ii} line at $8542.1\,\AA$ arrives at the formation height of the \ion{Ca}{II} K line core between $t=3.27\,\text{min}$ to $t=3.73\,\text{min}$. Therefore, we define $t_c = 3.27\,\text{min}$ as the transition time between the chromospheric drain phase and the subsequent upward-flow phase. This is the time for which we conjecture that the shock dissipated in the formation height of the \ion{Ca}{II} K line core. Changes in the acoustic power can give further clues on the shock dissipation, hence we analyze the changes in the spatial distribution of the acoustic power. We define the acoustic power excess as the difference between the integrated power one minute after $t_c$ and the integrated power one minute before $t_c$. Figure \ref{Fig7} summarises the results of this analysis for the $K_{2V}$ band (left panels) and for the line-core minimum (right panels) in the defined frequency ranges (medium-frequency range top panels and high-frequency range bottom panels). 

The magnitude of the acoustic power excess is more intense in the $K_{2V}$ band (ten times bigger) than in the line-core minimum for both frequency ranges, indicating the dominance of characteristic modes in the low chromosphere. In addition, the spatial distribution of the areas of positive or negative acoustic power excess differs, apparently being associated with structures previously identified in Sect.\,\ref{swirl}.

\begin{figure}
\centering
\includegraphics[width=9cm]{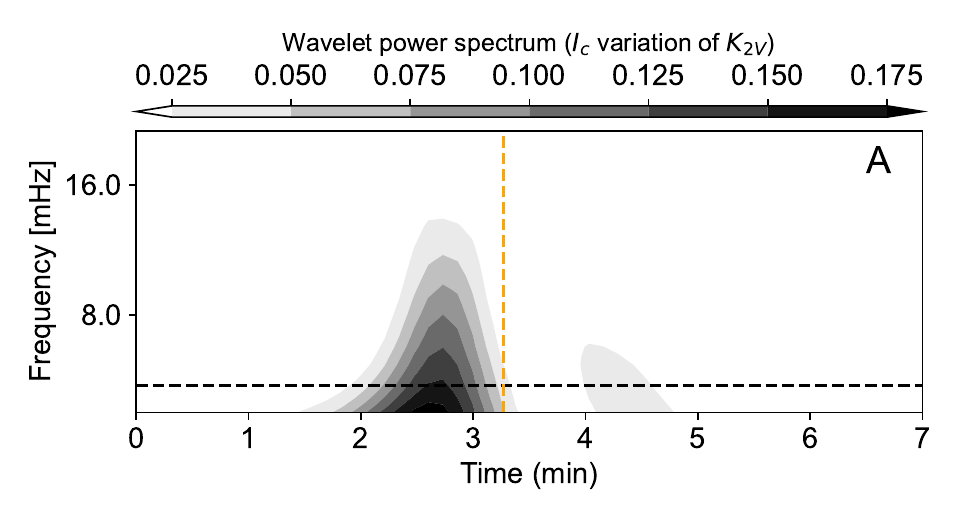}
\caption{Wavelet power-spectrum of a pixel marked by a black cross ($\times$) in the zoomed view of the upper left map of Fig.\,\ref{Fig7} and labelled as A. The horizontal black dashed line marks $5.5\,$mHz and the vertical orange dashed line refers to the time $t=3.27\,\text{min}$.}
\label{Fig7.1}
\end{figure}

\begin{figure}
\centering
\includegraphics[width=9cm]{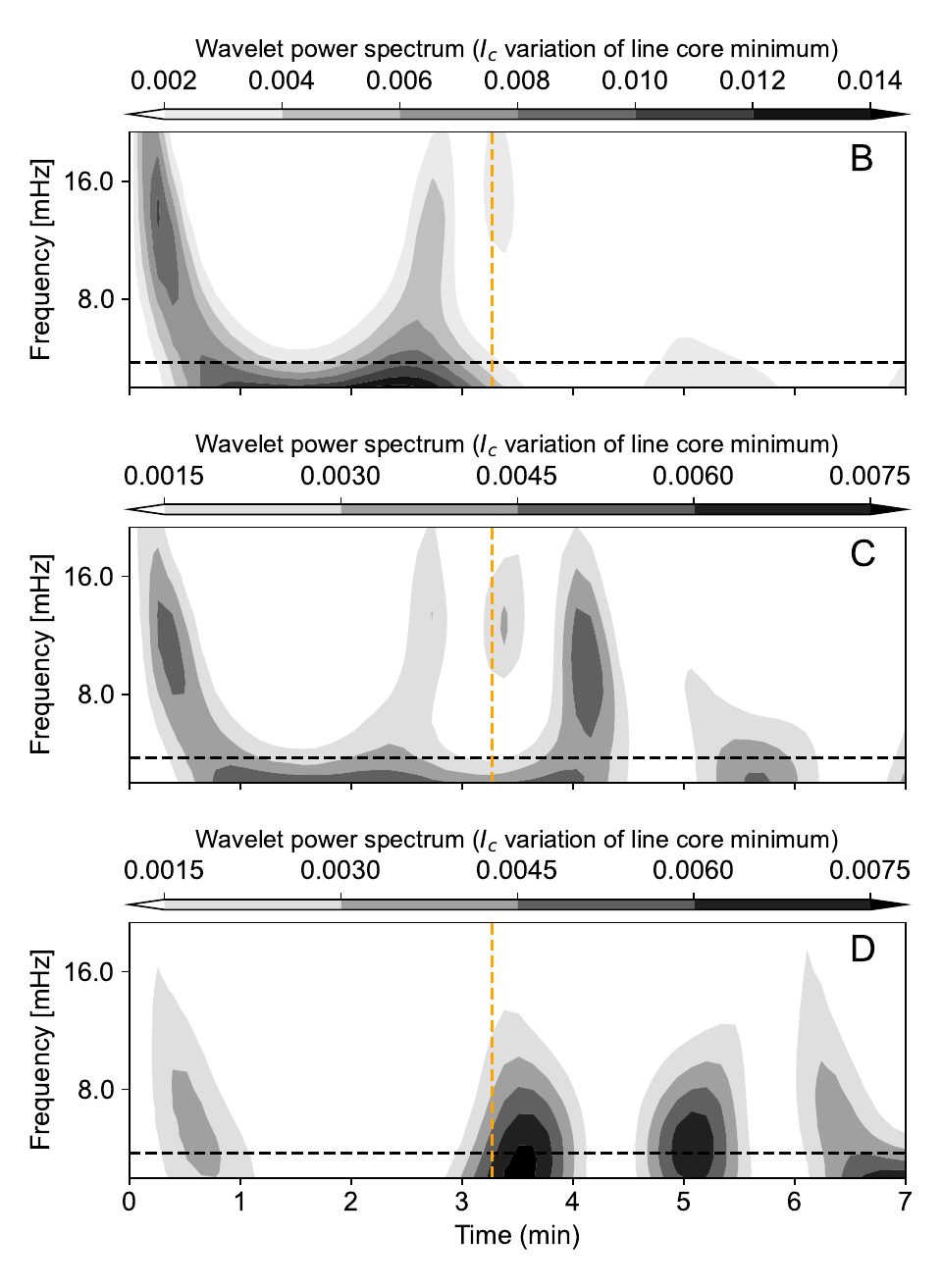}
\caption{Wavelet power-spectra of the three pixels marked by a red, magenta and cyan plus signs ($+$) in the zoomed inset in the top right map of Fig.\,\ref{Fig7} and labelled as B, C, and D respectively. The horizontal black dashed line marks $5.5\,$mHz and the vertical orange dashed line refers to the time $t=3.27\,\text{min}$.}
\label{Fig7.2}
\end{figure}

The negative power excess in the medium-frequency range, visible in the central region of the $K_{2V}$ power map, is associated with the chromospheric bright grain observed in the drain phase (see central red contour in the fourth row of Fig.\,\ref{Fig5}). This patch of negative excess corresponds to oscillations in the intensity of the $K_{2V}$ band above the cutoff frequency detected before $t_c$ which disappear thereafter. To clarify the meaning of a negative power excess, Fig.\,\ref{Fig7.1} shows a wavelet power spectrum of the intensity variations of the $K_{2V}$ band for a representative pixel within the negative power excess patch (black $\times$-sing in the zoomed inset in the upper left panel of Fig.\,\ref{Fig7}). This spectrum shows a power enhancement in the defined frequency ranges, especially around 5.5 mHz, but only before $t_c$ (dashed vertical orange line), which leads to the detected negative excess. Although no large accumulation of positive power excess is seen in the $K_{2V}$ power map, there are some faint patches localised in the right-hand side of the FOV, presumably associated with the chromospheric swirl. 

The intensity variation in the line-core minimum shows two small patches of negative power excess in different locations near the magnetic flux concentration. These patches of negative excess correspond to oscillations in the intensity of the line-core minimum above the cutoff frequency detected before $t_c$ which disappear thereafter. The intensity variations of the line-core minimum provide information on the maximum local opacity variation, which is mainly determined by changes in temperature stratification. Therefore these two patches may correspond to regions of wave dissipation. In contrast, patches of positive excess found in the FOV follow the spiral structure of the chromospheric swirl in the close vicinity of the MBP. 

Figure \ref{Fig7.2} depicts three wavelet power spectra of the line-core minimum for three pixels in the region of positive and negative power excess near the MBP. The top plot shows the power spectrum of pixel B (see pixel location on the zoomed inset of the top right panel of Fig.\,\ref{Fig7}) localised in a patch of negative power excess within the MBP area. Similar to Fig.\,\ref{Fig7.1}, the majority of the power occurs before $t_c$, albeit with a greater contribution from high-frequency modes. The middle and bottom plots show the power spectrum of pixels C and D respectively, which are localised in two distinct patches of positive power excess: one near the MBP area and the other one situated farther away from it. Both spectra show power enhancements after $t_c$ (dashed vertical orange line) but with a different frequency distribution. On the one hand, pixel C shows an excitation of modes above $8\,$mHz for 30 seconds after $t_c$, and then an excitation extended in frequency at around $t=4.0\,\text{min}$. On the other hand, pixel D shows power enhancements at frequencies close to $5.5\,$mHz between $t=3.0\,\text{min}$ and $t=4.0\,\text{min}$ which is then repeated at $t=5.0\,\text{min}$. The differences between these two pixels suggest that different physical processes take place in these two locations.

The power excesses in the high-frequency range show a similar pattern as the power excesses in the middle-frequency range but are less intense, due to the low amplitude of these modes. In particular, the negative power excess in the medium-frequency range, visible in the central region of the $K_{2V}$ power map has its counterpart in the high-frequency range. The wavelet power spectrum shown in Fig.\,\ref{Fig7.1} reveals high power in an extended range of frequencies which goes from the cutoff frequency up to and above $8\,$mHz. As proposed by \cite{Carlsson1997}, high-frequency oscillations associated with chromospheric bright grains can be produced by resonant interference between weak high-frequency modes and medium-frequency waves leading them to steepen rapidly. 

After the time $t_c$, the positive power excess in the medium-frequency range, visible in the central region of the map of the line-core minimum, has also a counterpart in the high-frequency range. The patch of positive power excess near the MBP shown in the bottom right panel of Fig.\,\ref{Fig7} is an indication that the energy dissipation can induce high-frequency oscillations at the formation height of the line-core minimum.

In a nutshell, we find power excess near the cutoff frequency at the location of the chromospheric bright grain in the chromospheric drain phase, which supports the shock wave hypothesis as such an oscillatory signature can be expected in connection with the formation of a shock wave. After the point in time that separates the drain phase from the upward-flow phase (which is also the arrival time of the shock front as detected in Ca II K), we find a lack of power or power enhancement at high frequencies, presumably due to dissipation effects and associated excitation of high-frequency waves, respectively.

\end{appendix}

\end{document}